\documentclass[a4paper,12pt]{article}
\usepackage{test,caption}

\usepackage{xcolor}
\definecolor{pantone}{RGB}{1, 33, 105}

\usepackage{setspace} 
\usepackage{hyperref}
\hypersetup{colorlinks,linkcolor=blue,citecolor=blue,urlcolor=pantone}
\usepackage{doi}

\usepackage[style=ext-authoryear-comp,
sorting=nyt, 
sortcites=false, 
dashed=false, 
maxcitenames=2, 
maxbibnames=99, 
uniquelist=false,
uniquename=false,
giveninits=true, 
natbib, 
date=year 
]{biblatex}

\AtBeginRefsection{\GenRefcontextData{sorting=ynt}}
\AtEveryCite{\localrefcontext[sorting=ynt]}

\DeclareFieldFormat{pages}{#1} 
\renewbibmacro{in:}{\ifentrytype{article}{}{\printtext{\bibstring{in}\intitlepunct}}} 

\DeclareFieldFormat[article,inbook,incollection,inproceedings,patent,thesis,unpublished]{titlecase:title}{\MakeSentenceCase*{#1}} 

\newbibmacro*{atoda:titlelink}[1]{%
  \ifhyperref
    {\iffieldundef{doi}
       {\iffieldundef{eprint}
          {\iffieldundef{url}
             {#1}
             {\href{\thefield{url}}{#1}}}
          {\href{https://arxiv.org/abs/\thefield{eprint}}{#1}}}
       {\href{https://doi.org/\thefield{doi}}{#1}}}
    {#1}}

\DeclareFieldFormat{title}{\usebibmacro{atoda:titlelink}{\mkbibemph{#1}}}
\DeclareFieldFormat
  [article,inbook,incollection,inproceedings,patent,thesis,unpublished]
  {title}{\usebibmacro{atoda:titlelink}{\mkbibquote{#1\isdot}}}

\renewbibmacro*{doi+eprint+url}{}

\newcommand{\cI}{\mathcal{I}}
\newcommand{\cJ}{\mathcal{J}}
\newcommand{\cL}{\mathcal{L}}

\usepackage[inline,shortlabels]{enumitem}
\setlist[enumerate,1]{label=(\roman*)}

\usepackage[margin = 1.25in]{geometry}
\title{Leverage, Endogenous Unbalanced Growth, and Asset Price Bubbles\thanks{This research was financially supported by Japan Center for Economic Research; Japan Securities Scholarship Foundation; the Joint Usage/Research Center, Institute of Economic Research, Hitotsubashi University (Grant ID: IERPK2515); JSPS KAKENHI (Grant Number JP24K00269). We thank Fumio Hayashi, Keiichi Kishi, 
Jos\'e Scheinkman, Joseph Stiglitz, and seminar participants at various institutions and conferences for comments and encouragement.}
}
\author{Tomohiro Hirano\thanks{Department of Economics, Royal Holloway, University of London. Email: \href{mailto:tomohiro.hirano@rhul.ac.uk}{tomohiro.hirano@rhul.ac.uk}.} \and Ryo Jinnai\thanks{Institute of Economic Research, Hitotsubashi University. Email: \href{mailto:ryo.jinnai@r.hit-u.ac.jp}{ryo.jinnai@r.hit-u.ac.jp}.} \and Alexis Akira Toda\thanks{Department of Economics, Emory University and Research Institute for Economics and Business Administration, Kobe University. Email: \href{mailto:alexis.akira.toda@emory.edu}{alexis.akira.toda@emory.edu}.}}

\numberwithin{equation}{section}

\begin{document}

\maketitle

\begin{abstract}
We develop a macro-finance model in which leverage creates a positive feedback loop between capital investment and land prices. When leverage is below a threshold, land prices equal the present value of rents. Relaxing leverage lowers the productivity of the marginal investor and the interest rate until the fundamental value diverges. The economy then undergoes a phase transition to unbalanced growth. Demand for a store of value makes land prices grow faster than rents, so a bubble necessarily emerges. When the upper tail of the productivity distribution is sufficiently thick, this regime can prevail at arbitrarily high leverage.

\medskip

\noindent
\textbf{Keywords:} bubble necessity, financial accelerator, leverage, phase transition, rational bubbles, unbalanced growth.

\medskip
\noindent
\textbf{JEL codes:} D53, E44, G12, O41.
\end{abstract}

\section{Introduction}

Economic historians have long linked booms in asset prices to expansions in credit. \citet[Ch.~4]{Kindleberger2005} state, ``Axiom number two. Asset price bubbles depend on the growth of credit. \dots every mania has been associated with the expansion of credit.'' Formal empirical evidence points in the same direction. \citet{RajanRamcharan2015} find that greater credit availability inflated farmland prices during the U.S. boom of the 1920s. Using U.S. branching deregulations between 1994 and 2005, \citet{FavaraImbs2015} find that exogenous expansions in mortgage credit raised house prices. These findings motivate our theoretical question: can relaxing borrowing constraints affect not only the level of asset prices but also whether a dividend-paying asset trades above its fundamental value, defined as the present value of its dividends?

Since the 2008--2009 global financial crisis, economists have increasingly used rational bubble models (models with rational expectations and common beliefs in which an asset price exceeds its fundamental value) to study how financial frictions can sustain asset overvaluation. Most of this literature asks whether financial constraints can support a bubbly equilibrium (possibility), rather than whether a bubbly equilibrium must arise (necessity). In models with pure bubbles (intrinsically worthless assets like fiat money), a bubbleless equilibrium remains available under the same primitives, so showing that a bubble can arise does not show that it must arise. Moreover, many of these models imply that bubbly equilibria disappear once borrowing constraints become sufficiently loose.\footnote{\label{fn:RB}Examples include \citet{ArceLopezSalido2011,FarhiTirole2012,Kocherlakota2013,AokiNikolov2015,HiranoInabaYanagawa2015,HiranoYanagawa2017,Guerron-QuintanaHiranoJinnai2023}.} This prediction contrasts with the historical narrative that easier credit accompanies booms in asset prices. In this paper, we ask whether relaxing leverage can instead eliminate fundamental valuation and make a bubble necessary.

We develop a dynamic macro-finance model in which agents with heterogeneous investment productivity allocate saving between capital and land. Agents observe their productivity before choosing their portfolios and may finance capital investment up to a common leverage limit. Land is in fixed supply, pays positive rents, and serves as a safe store of value. To clarify the mechanism, we first solve a tractable model with a linear capital technology and an unmodeled financial sector that accommodates agents' net bond position at finite dates, while its own position becomes negligible relative to aggregate wealth in the long run. We characterize the financial conditions for balanced growth, where the economy and dividends grow at the same rate, and for unbalanced growth, where the economy grows faster than dividends. The model links land prices and capital investment through aggregate wealth. Agents with high productivity borrow and invest, while agents with low productivity hold safe claims. A higher land price raises wealth and supports more investment. Greater capital accumulation raises future wealth and the demand for land, which feeds back into its price. Leverage determines the strength of this feedback (financial accelerator). Productive capital does not absorb all saving because agents with low current productivity continue to demand a safe asset.

For each leverage level below a critical value, a unique aggregate capital level supports a balanced-growth equilibrium. Aggregate wealth has gross growth rate one, the gross interest rate exceeds one, and the land price equals the present value of rents. As leverage rises along this branch, productive agents borrow more and their wealth grows faster. Because aggregate wealth must continue to grow at gross rate one, the wealth of unproductive agents must grow more slowly, which requires a lower interest rate on safe claims. The gross interest rate falls toward one, but a fundamental equilibrium cannot have an interest rate below one: with positive constant rents, the present value of rents would be infinite, which is incompatible with a finite equilibrium land price. Thus, once leverage crosses the threshold, a fundamental equilibrium ceases to exist. Theorem~\ref{thm:trend} shows that the equilibrium must instead exhibit unbalanced growth. Capital, wealth, and land prices then grow at a common gross rate above one, while rents remain constant. Low-productivity agents nevertheless continue to demand safe claims. The financial sector can accommodate agents' aggregate net bond position at each finite date, but this position must become negligible relative to aggregate wealth. It therefore cannot absorb net demand for safe claims on the scale of the growing economy in the long run. Because land is the only safe asset in positive net supply, its fixed supply forces its price to grow with wealth. Constant rents cannot account for this growing price, so land necessarily contains a bubble. This is not merely one bubbly equilibrium among several: the post-threshold equilibrium is unique and bubbly whenever it exists. We call the change between these two regimes a \emph{phase transition}.\footnote{``Phase transition'' is a term from the natural sciences for a qualitative change in a system's state as a parameter crosses a critical value. For example, at fixed pressure, water changes from ice to liquid as temperature rises above the melting point. The analogy is appropriate here because crossing the leverage threshold changes the economy from balanced growth with fundamental land values to unbalanced growth with a land price bubble.} In the unbalanced-growth regime, the interest rate equals the economic growth rate and rises with leverage. The relation between leverage and the interest rate is therefore V-shaped: the interest rate falls toward one on the balanced-growth side and rises from one on the unbalanced-growth side.

Theorem~\ref{thm:trend} establishes bubble necessity whenever the post-threshold equilibrium exists. Proposition~\ref{prop:admissible_leverage} asks how far this bubble necessity regime extends as leverage rises. Why can it survive under loose financial conditions, whereas many existing models predict that bubbles disappear (footnote \ref{fn:RB})? The key is the productivity distribution. Many of these models assume only two productivity types, so their distributions have bounded support. As leverage rises, the productivity cutoff above which agents invest also rises. With bounded productivity or a sufficiently thin upper tail, average productivity above the cutoff eventually becomes too low relative to the cutoff to sustain unbalanced growth, so the post-threshold equilibrium ceases to exist. With a sufficiently thick unbounded upper tail, highly productive agents continue to generate enough investment demand for the bubble necessity regime to prevail at arbitrarily high leverage.

In \S\ref{sec:general}, we remove the linear technology, the unmodeled financial sector, and the assumption of an exogenous rent. At low leverage, the model has a unique steady state. At high leverage, it has no stationary equilibrium, and under additional conditions we construct a locally unique unbalanced-growth equilibrium for sufficiently large initial capital. Land rents are endogenous in this equilibrium, but they grow more slowly than land prices and therefore cannot account for the price. The land price contains a bubble. Our numerical exercise shows how capital, land prices, rents, and the price--rent ratio respond when leverage crosses the threshold.

\subsection{Related Literature}\label{sec:literature}

\paragraph{Rational bubbles}
Our paper belongs to the rational bubble literature, which studies whether an asset price can exceed the present value of its dividends under rational expectations and common beliefs. Foundational contributions establish both the possibility of such bubbles and the restrictions that equilibrium places on them \citep{Samuelson1958,Bewley1980,Tirole1985,ScheinkmanWeiss1986,Kocherlakota1992,SantosWoodford1997}. See \citet{MartinVentura2018,HiranoToda2024JME,Barlevy2025BubbleBook} for surveys. This literature has largely focused on pure bubbles, \ie, intrinsically worthless assets such as fiat money, in part because bubbles attached to positive dividends face stringent restrictions \citep[Theorem~3.3 and Corollary~3.4]{SantosWoodford1997}. Pure-bubble models also often admit a continuum of equilibria and do not yield a price--dividend ratio. By contrast, we study a dividend-paying asset in a model with a unique equilibrium, so the land price can be compared directly with the present value of rents.

\paragraph{Financial accelerator}

Our mechanism is related to the financial accelerator literature, in which balance sheets and borrowing constraints amplify shocks and asset-price movements \citep{KiyotakiMoore1997,BernankeGertlerGilchrist1999,BrunnermeierSannikov2014}. The canonical models in this literature do not ask whether the asset price violates the no-bubble condition. In our model, a higher land price raises entrepreneurial net worth and investment; capital accumulation in turn raises future wealth, land rents, and the demand for land. When leverage makes this feedback sufficiently strong, it changes not merely the magnitude of the response but the economy's long-run growth and asset-pricing regime.

\paragraph{Rational bubbles and financial conditions}
An extensive literature studies how rational bubbles interact with financial constraints, credit, and investment \citep{CaballeroKrishnamurthy2006,FarhiTirole2012,MartinVentura2012,HiranoInabaYanagawa2015}. Three papers are especially close to ours. \citet{MartinVentura2016} introduce pure bubbles into a Diamond--Tirole overlapping-generations economy and study how the resulting bubbly collateral affects credit and investment; at fixed primitives, bubble size varies across rational-expectations equilibria. \citet{HiranoYanagawa2017} show that pure bubbles arise for an intermediate range of pledgeability: sufficiently tight constraints cannot sustain a bubble, whereas sufficiently loose constraints raise the interest rate enough to eliminate it. \citet{BiswasHansonPhan2020} adopt a related heterogeneous-productivity environment with Pareto productivity and study the consequences of a stochastic bubble collapse in the presence of downward wage rigidity. These are possibility results: they identify or study pure-bubble equilibria under specified financial conditions, but the existence of a bubbly equilibrium does not rule out a bubbleless equilibrium. Our result is a necessity result. We vary leverage across unique equilibria and show that relaxing a constraint that continues to bind can eliminate the fundamental balanced-growth equilibrium. Whenever the post-threshold equilibrium exists, it exhibits unbalanced growth and land is necessarily bubbly. When productivity has a sufficiently thick unbounded upper tail, this bubble necessity regime can prevail at arbitrarily high finite leverage. Positive land rents make valuation determinate. We provide a detailed comparison with \citet{HiranoYanagawa2017,BiswasHansonPhan2020} in \S\ref{sec:loose_constraints}.

\paragraph{Bubble necessity}
Recent research on rational bubbles has reframed the central question from whether they \emph{can} occur (possibility) to whether they \emph{must} occur (necessity).
\citet{HiranoToda2025JPE} prove the Bubble Necessity Theorem: if the asymptotic dividend growth rate lies between the counterfactual bubbleless interest rate and the economic growth rate, every equilibrium is bubbly.\footnote{An early antecedent is \citet[\S7]{Wilson1981}, who provides an example in which no fundamental equilibrium exists. \citet[Proposition~1(c)]{Tirole1985} argues that bubbles are necessary when the counterfactual bubbleless interest rate is negative, but \citet{PhamToda2026ECMA} construct a counterexample and correct this claim.}
\citet{HiranoToda2025PNAS} develop this approach for land under aggregate uncertainty and show how unbalanced growth can make land overvaluation inevitable. \citet{Sorger2026} establishes the necessity of stock price bubbles in simple growth models with positive profits, where wages and dividends grow at different rates.
\citet{HiranoToda2025JPE} take dividends as exogenous, while \citet{HiranoToda2025PNAS} endogenize land rents; neither paper studies how leverage selects between balanced- and unbalanced-growth regimes or generates a phase transition. The present paper endogenizes this regime switch through leverage.

\paragraph{Related notions of bubbles}
Other theories use the term ``bubble'' for mechanisms distinct from rational bubbles under common beliefs. With heterogeneous beliefs and short-sale or collateral constraints, investors may pay a speculative premium because they expect to resell the asset to a more optimistic investor \citep{HarrisonKreps1978,ScheinkmanXiong2003,FostelGeanakoplos2012Tranching}. Because valuations differ across investors, these models are conceptually distinct from the common-belief rational bubbles studied here. Agency and risk shifting provide another mechanism through which credit expansion can generate asset-price booms \citep{AllenGale2000,AllenBarlevyGale2022}. Finally, \citet{MiaoWang2018} describe self-fulfilling fluctuations in firm value and collateral as rational stock price bubbles. However, as \citet{HiranoToda2025EJW} prove, the stock price in the Miao--Wang model equals the present value of dividends in each equilibrium. Under the $P>V$ definition used here, their mechanism is therefore more appropriately viewed as multiplicity among fundamental equilibria selected by self-fulfilling beliefs. Our result instead concerns when a unique equilibrium price necessarily exceeds the present value of rents.

\section{Tractable \texorpdfstring{$AK$}{AK} benchmark}\label{sec:simple}

To develop intuition and obtain a closed-form solution, we present a simple model with two features. First, capital investment uses a linear $AK$ technology. Second, agents can trade a one-period bond with an unmodeled financial sector at a constant gross risk-free rate. The sector accommodates agents' aggregate net bond position at every finite date, but that position must become negligible relative to aggregate wealth in the long run. Thus the benchmark is almost closed in the long run, and this condition determines the interest rate. These assumptions make the benchmark analytically tractable. \S\ref{sec:general} shows that the mechanism survives under more general assumptions.

\subsection{Model}

\paragraph{Agents and timing}
The economy is populated by a continuum of agents with mass 1 indexed by $i\in I=[0,1]$.\footnote{A continuum of independent random variables indexed by the Lebesgue unit interval raises a measurability issue. We invoke the Fubini-extension construction of \citet{SunZhang2009}, which delivers an exact law of large numbers and justifies the aggregation below. Alternatively, one may interpret aggregate quantities as limits of per capita quantities in finite-agent economies.}
A typical agent has logarithmic utility
\begin{equation}
\E_0\sum_{t=0}^\infty \beta^t \log c_{it},\label{eq:log_utility}
\end{equation}
where $\beta\in (0,1)$ is the discount factor and $c_{it}>0$ is consumption.
At the beginning of period $t$, agent $i$ enters with wealth $w_{it}>0$ carried over from the previous period and observes investment productivity $z_{it}\ge 0$. The agent then chooses consumption, capital investment, and positions in land and a one-period bond.

\paragraph{Investment technology and productivity}

Investment productivity determines how efficiently current goods are transformed into capital. If agent $i$ invests $i_{it}\ge0$ units of the good, the investment produces $k_{i,t+1}=z_{it}i_{it}$ units of capital at date $t+1$. Each unit of capital produces $m>0$ units of the good, so the investment generates income $mz_{it}i_{it}$. Although we could combine $m$ and $z_{it}$ under linear production, keeping them separate facilitates comparison with the general model in \S\ref{sec:general}. Following \citet{Woodford1990,Kiyotaki1998,HiranoYanagawa2017}, agents observe heterogeneous investment opportunities before choosing investment. This ex ante heterogeneity creates gains from reallocating funds from low- to high-productivity agents and makes leverage economically relevant.

The economy features no aggregate uncertainty but agents are subject to idiosyncratic risk. As a benchmark, we rule out sunspots and focus on deterministic aggregate equilibrium sequences. Productivity $z_{it}$ is independently and identically distributed across agents and over time with cumulative distribution function (cdf) $\Phi$ satisfying the following assumption.

\begin{asmp}\label{asmp:Phi}
$\Phi:[0,\infty)\to [0,1]$ is absolutely continuous with $\int_0^\infty z\diff \Phi(z)<\infty$.
\end{asmp}

We adopt log utility and the \iid specification for analytical tractability.\footnote{An earlier draft of this paper, \citet{HiranoJinnaiTodaBubble}, discusses the cases of power utility (Appendix C.1) and Markovian shocks (Appendix C.2). The example in \citet[\S V.B]{HiranoToda2025JPE}, which builds on the present model, abstracts from borrowing but considers a Markovian environment.} 
The absolute continuity of $\Phi$ implies that the productivity distribution has no point mass except possibly at $z=0$. This assumption is inessential but simplifies the analysis by avoiding cases. The condition $\int_0^\infty z\diff \Phi(z)<\infty$ implies that the mean productivity is finite, which is necessary to ensure that aggregate capital is finite.

\paragraph{Land, bonds, and budget constraints}

There is a unit supply of land, which pays a constant dividend (rent) $D>0$ each period. Let $P_t$ be its ex-dividend price and let $\theta_{it}$ be agent $i$'s land holdings. Agents can also trade a one-period bond with an unmodeled financial sector at a constant gross risk-free rate $R>1$, determined in equilibrium below. The sector accommodates agents' aggregate net bond position at that rate by taking the opposite position.\footnote{\label{fn:financial_sector}The financial sector may be interpreted as a central bank or another public liquidity facility, a financial intermediary with sufficient balance-sheet capacity, or the rest of the world.} Let $n_{it}$ denote the market value of agent $i$'s net bond position with the financial sector; $n_{it}>0$ represents lending and $n_{it}<0$ borrowing. At the beginning of date 0, agent $i$ owns $k_{i0}\ge0$ units of capital and $\theta_{i,-1}\ge0$ units of land and has no initial bond position, $n_{i,-1}=0$. We assume $k_{i0}+\theta_{i,-1}>0$ for all $i$,
\begin{equation}
    \int_I k_{i0}\diff i=K_0>0,
    \qquad
    \int_I\theta_{i,-1}\diff i=1, \label{eq:K0}
\end{equation}
and that date-0 productivity is independent of these predetermined holdings. For $t\ge0$, the budget and next-period wealth equations, together with the initial wealth condition, are
\begin{subequations}\label{eq:budget}
\begin{align}
    c_{it}+i_{it}+P_t\theta_{it}+n_{it}&=w_{it}, \label{eq:budget_t}\\
    w_{i,t+1}&=mz_{it}i_{it}+(P_{t+1}+D)\theta_{it}+Rn_{it}, \label{eq:budget_t+}\\
    w_{i0}&=mk_{i0}+(P_0+D)\theta_{i,-1}. \label{eq:initial_wealth_individual}
\end{align}
\end{subequations}
Thus initial capital produces $mk_{i0}$ units of date-0 income, while initial land pays its date-0 rent and can be sold at $P_0$. Letting $W_t=\int_I w_{it}\diff i$ be the aggregate wealth, \eqref{eq:initial_wealth_individual} and \eqref{eq:K0} imply $W_0=mK_0+P_0+D$.
Define the value of the agent's total risk-free asset position by
\begin{equation}
    a_{it}\coloneqq P_t\theta_{it}+n_{it}. \label{eq:riskfree_position}
\end{equation}
Land and the bond are perfect substitutes in individual portfolios. Indeed, the zero-cost perturbation $(\theta_{it},n_{it})\mapsto(\theta_{it}+\epsilon,n_{it}-P_t\epsilon)$ leaves the budget constraint, $a_{it}$, and the leverage constraint in \eqref{eq:leverage} below unchanged. Absence of arbitrage therefore requires
\begin{equation}
    P_{t+1}+D=RP_t. \label{eq:noarbitrage_R}
\end{equation}
Using \eqref{eq:riskfree_position} and \eqref{eq:noarbitrage_R}, next-period wealth becomes
\begin{equation}
    w_{i,t+1}=\underbrace{mz_{it}i_{it}}_\text{income from capital}+\underbrace{Ra_{it}}_\text{income from safe claims}. \label{eq:budget_t+_reduced}
\end{equation}

\paragraph{Leverage constraint}

Agents are subject to the leverage constraint
\begin{equation}
    0\le \underbrace{i_{it}}_\text{investment}\le \lambda(\underbrace{i_{it}+a_{it}}_\text{equity}), \label{eq:leverage}
\end{equation}
where $\lambda\ge 1$ is the exogenous leverage limit. Here $i_{it}+a_{it}=w_{it}-c_{it}$ is the agent's equity after consumption. We interpret \eqref{eq:leverage} as a contemporaneous maximum-leverage restriction imposed after productivity is observed, rather than as a state-contingent solvency constraint. The constraint is a reduced-form implication of limited enforcement or limited pledgeability of capital investment: capital investment cannot exceed a multiple of total equity.\footnote{According to standard balance-sheet accounting, equity equals assets ($i_{it}$) minus liabilities ($-a_{it}$), which is $i_{it}+a_{it}$. One limited-enforcement interpretation is that an agent can divert $i_{it}/\lambda$ units of capital investment and forfeits equity if diversion is detected. The no-diversion condition $i_{it}/\lambda\le i_{it}+a_{it}$ is exactly \eqref{eq:leverage}. The leverage constraint is identical to that in \citet{BueraShin2013}; see their footnote 9 or \citet{BanerjeeNewman1991} for related microfoundations.} Importantly, land does not provide a separate collateral service in this benchmark; only the total value $a_{it}$ of safe claims enters the constraint. Since $i_{it}\ge 0$ and $\lambda\ge 1$, equity is nonnegative. Equivalently,
\begin{equation}
    \underbrace{-a_{it}}_\text{net borrowing} \le \underbrace{(1-1/\lambda)}_\text{maximum debt-to-investment ratio}i_{it}. \label{eq:borrowing_constraint}
\end{equation}

\paragraph{Individual behavior}

Given initial wealth, each agent maximizes the lifetime utility \eqref{eq:log_utility} subject to the budget constraints \eqref{eq:budget} and the leverage constraint \eqref{eq:leverage}. Log utility implies the optimal consumption rule $c_{it}=(1-\beta)w_{it}$ (see, for instance, \citet[Example 15.2]{TodaEME}).

For an agent with productivity $z$, the gross return on investment is $mz$, whereas the return on safe claims is $R$. Therefore, the agent invests if and only if $mz>R$, or $z>\bar{z}\coloneqq R/m$. Whenever $z>\bar{z}$, maximal leverage is optimal. Hence
\begin{equation}
    (i_{it},a_{it})=\begin{cases*}
        (0,\beta w_{it}) & if $z_{it}\le \bar{z}$,\\
        (\lambda \beta w_{it},(1-\lambda)\beta w_{it}) & if $z_{it}>\bar{z}$.
    \end{cases*}\label{eq:asset_allocation}
\end{equation}

\subsection{Equilibrium}

Using \eqref{eq:budget_t+_reduced} and \eqref{eq:asset_allocation}, individual wealth evolves according to
\begin{equation}
    w_{i,t+1}=\beta(R+\lambda\max\set{0,mz_{it}-R})w_{it}. \label{eq:wit}
\end{equation}
Aggregating across agents gives
\begin{equation}
    W_{t+1}=\beta(R+\lambda\pi(R))W_t, \label{eq:Wt1}
\end{equation}
where
\begin{equation}
    \pi(R)\coloneqq \int_0^\infty\max\set{0,mz-R}\diff \Phi(z)\label{eq:pi}
\end{equation}
is the expected excess return on unlevered capital investment. The aggregate wealth dynamics \eqref{eq:Wt1} implies that aggregate wealth grows at a constant gross growth rate
\begin{equation}
    G\coloneqq \beta(R+\lambda\pi(R)). \label{eq:G}
\end{equation}
Aggregating individual capital $k_{i,t+1}=z_{it}i_{it}$ across agents and noting that productivities are independent across agents, we obtain
\begin{equation}
    K_{t+1}=\beta\lambda W_t\int_{\bar{z}}^\infty z\diff \Phi(z). \label{eq:Kt+}
\end{equation}
Land-market clearing and the aggregate net bond position are
\begin{equation*}
    \int_I\theta_{it}\diff i=1,
    \qquad
    N_t\coloneqq\int_I n_{it}\diff i.
\end{equation*}
Therefore, aggregating $a_{it}=P_t\theta_{it}+n_{it}$ in \eqref{eq:asset_allocation} yields
\begin{align}
    P_t+N_t&=\underbrace{\beta W_t\int_0^{\bar{z}}\diff \Phi(z)}_\text{savings by unproductive agents} +\underbrace{\beta(1-\lambda)W_t\int_{\bar{z}}^\infty \diff \Phi(z)}_\text{borrowing by productive agents}\notag \\
    &=\beta (\lambda\Phi(\bar{z})+1-\lambda)W_t. \label{eq:Pt1}
\end{align}
We call a productivity threshold $\bar{z}$ \emph{admissible} if
\begin{equation*}
    \lambda\Phi(\bar{z})+1-\lambda>0,
\end{equation*}
or equivalently $\Phi(\bar{z})>1-1/\lambda$. This condition ensures that aggregate net demand for safe claims in \eqref{eq:Pt1} is positive. If $\Phi$ is strictly increasing, it is also equivalent to $\bar{z}>\Phi^{-1}(1-1/\lambda)$. We define an equilibrium as follows.

\begin{defn}\label{defn:lr}
Given initial aggregate capital $K_0>0$, a \emph{rational expectations equilibrium} consists of a gross risk-free rate $R>1$, a productivity threshold $\bar{z}>0$, an economic growth rate $G>0$, and a sequence $\set{(W_t,K_t,P_t,N_t)}_{t=0}^\infty$, with $W_t,K_t,P_t>0$ and $N_t\in\R$, satisfying the initial condition $W_0=mK_0+P_0+D$ and the following conditions:
\begin{enumerate*}
    \item $R=(P_{t+1}+D)/P_t$,
    \item\label{item:zbarlambda} $\bar{z}$ is admissible and satisfies $\bar{z}=R/m$,
    \item $G$ satisfies \eqref{eq:G},
    \item $W_t$ satisfies \eqref{eq:Wt1},
    \item $K_{t+1}$ satisfies \eqref{eq:Kt+},
    \item $(P_t,N_t)$ satisfies \eqref{eq:Pt1}, and
    \item\label{item:N/W} $N_t/W_t\to 0$ as $t\to\infty$.
\end{enumerate*}
\end{defn}

We impose condition \ref{item:N/W} to make the benchmark almost closed in the long run.\footnote{One deliberately stylized interpretation is that an institution of the type described in footnote~\ref{fn:financial_sector} selects a constant interest rate $R$ so that the position it accommodates becomes negligible relative to aggregate wealth in the long run, $N_t/W_t\to0$.} If risk-free claims cleared exclusively among agents, their aggregate net position would satisfy $N_t=0$ at every date. The benchmark instead permits $N_t\ne0$ at finite dates, which makes a constant-interest-rate solution possible. The financial sector takes the opposite position $-N_t$, and, given $R$, equation \eqref{eq:Pt1} determines $N_t$ as the residual after agents make their date-$t$ portfolio choices. The condition $N_t/W_t\to0$ ensures that the sector can accommodate these finite-date positions without remaining a net source or sink for a nonvanishing share of aggregate wealth. This long-run market-clearing requirement determines $R$.

Although Definition \ref{defn:lr} allows $G>0$, Lemma \ref{lem:G<1} in the proof of Theorem \ref{thm:trend} shows that every equilibrium satisfies $G\ge1$. There are therefore only two cases. We say that an equilibrium exhibits \emph{balanced growth} if $G=1$ and \emph{unbalanced growth} if $G>1$. 
The terminology follows from the behavior of aggregate economic activity relative to dividends. 
The constant dividend $D>0$ has gross growth rate 1. Hence $G=1$ is balanced growth because the economy (aggregate wealth) and dividends grow at the same rate, while $G>1$ is unbalanced growth because the economy grows faster than dividends.

To study the asset pricing implications of our models, we briefly review the standard definition of a rational asset price bubble and a useful characterization. We restrict attention to deterministic price--dividend sequences, consistent with our exclusion of sunspots. Consider an infinitely-lived asset with ex-dividend price $P_t>0$ and dividend $D_t\ge0$. Let $q_t>0$ denote the date-0 price of a zero-coupon bond with maturity $t$, normalized by $q_0=1$. No arbitrage requires $q_tP_t=q_{t+1}(P_{t+1}+D_{t+1})$. Iterating this condition forward gives
\begin{equation}
    P_t
    =\underbrace{\frac{1}{q_t}\sum_{s=t+1}^\infty q_sD_s}_{\text{fundamental value $V_t$}}
    +\underbrace{\frac{1}{q_t}\lim_{T\to\infty}q_TP_T}_{\text{bubble component $B_t$}}. \label{eq:bubble_decomposition}
\end{equation}
Thus there is no bubble ($B_t=0$ for all $t$) if and only $\lim_{T\to\infty}q_TP_T=0$, which we call the \emph{no-bubble condition}. We call the asset price \emph{fundamental} if $P_t=V_t$ at every date and \emph{bubbly} if $P_t>V_t$ at every date. The Bubble Characterization Lemma of \citet[Proposition~7]{Montrucchio2004} and \citet[Lemma~1]{HiranoToda2025JPE} states that there is a bubble ($B_t>0$ for all $t$) if and only if
\begin{equation}
    \sum_{t=1}^\infty\frac{D_t}{P_t}<\infty. \label{eq:bubble_characterization}
\end{equation}
In the present model, $q_t=R^{-t}$ and $D_t=D$, so $V_t=D/(R-1)$; we classify equilibria accordingly. For more details, see \citet{HiranoToda2024JME,HiranoToda2025JPE,HiranoToda2025EJW}.

Returning to our model, define the return--growth gap
\begin{equation}
    \Psi(v,\lambda)\coloneqq (1-\beta)v-\beta\lambda\int_v^\infty (z-v)\diff\Phi(z). \label{eq:Psi}
\end{equation}
Here $v>0$ is a generic productivity threshold; when applied to the present model, $v=\bar z=R/m$. Setting $R=mv$ in \eqref{eq:G}, we obtain
\begin{equation}
    G-R=\beta\left(R+\lambda\int_0^\infty\max\set{0,mz-R}\diff\Phi(z)\right)-R=-m\Psi(v,\lambda). \label{eq:return_growth_gap}
\end{equation}
Thus $\Psi(v,\lambda)<0$ means that economic growth exceeds the risk-free return, whereas $\Psi(v,\lambda)>0$ means the opposite. The following condition ensures that the leverage threshold derived below is well defined and admissible.

\begin{asmp}\label{asmp:bmz}
Let $Z$ be a random variable with cdf $\Phi$. The model parameters satisfy
\begin{equation}
    \beta\E[Z\mid Z\ge 1/m]>1/m .\label{eq:bmz}
\end{equation}
\end{asmp}

Assumption \ref{asmp:bmz} is equivalent to $\beta\E[mZ\mid mZ\ge1]>1$. Because $\beta$ is the propensity to save and $mZ$ is the gross return on capital, when the safe gross return is one, the average wealth of agents whose productivity is high enough to invest can grow, making an economic takeoff possible. The following lemma collects the properties of the return--growth gap that locate the transition.

\begin{lem}[Leverage threshold]\label{lem:leverage_threshold}
Suppose Assumptions \ref{asmp:Phi} and \ref{asmp:bmz} hold. For every $\lambda>0$, $\Psi(\cdot,\lambda)$ is strictly increasing and has a unique positive root $v_\lambda$, characterized by $\Psi(v_\lambda,\lambda)=0$. The root $v_\lambda$ is strictly increasing in $\lambda$. Moreover, the leverage threshold
\begin{equation}
    \bar{\lambda}\coloneqq \frac{1-\beta}{\beta}\frac{1}{\int_{1/m}^\infty (mz-1)\diff \Phi(z)}, \label{eq:lambda_threshold}
\end{equation}
is positive and finite, and it satisfies
\begin{equation}
    \lambda\gtreqless\bar{\lambda}
    \iff \Psi(1/m,\lambda)\lesseqgtr 0
    \iff v_\lambda\gtreqless 1/m. \label{eq:Psi_threshold}
\end{equation}
Furthermore, the pair $(\bar{\lambda},1/m)$ is admissible.
\end{lem}

\begin{proof}
Differentiating \eqref{eq:Psi} gives
\begin{equation}
    \Psi_v(v,\lambda)=1-\beta+\beta\lambda[1-\Phi(v)]>0. \label{eq:Psi_v}
\end{equation}
Assumption \ref{asmp:bmz} implies $\E[Z]>0$, so $\Psi(0,\lambda)=-\beta\lambda\E[Z]<0$. Assumption \ref{asmp:Phi} implies $\int_v^\infty(z-v)\diff\Phi(z)\to0$, so $\Psi(v,\lambda)>0$ for all sufficiently large $v$. Hence the root $v_\lambda$ exists and is unique. At the root, $\Psi_\lambda(v_\lambda,\lambda)=-\beta\int_{v_\lambda}^\infty(z-v_\lambda)\diff\Phi(z)<0$, so the implicit function theorem and \eqref{eq:Psi_v} imply that $v_\lambda$ is strictly increasing in $\lambda$. Assumption \ref{asmp:bmz} also makes the denominator in \eqref{eq:lambda_threshold} positive. Substitution gives $\Psi(1/m,\bar\lambda)=0$, and the monotonicity of $\Psi$ in $v$ and $\lambda$ yields \eqref{eq:Psi_threshold}. Finally, substituting \eqref{eq:lambda_threshold} shows that the admissibility condition $\bar\lambda[1-\Phi(1/m)]<1$ is equivalent to \eqref{eq:bmz}.
\end{proof}

The lemma identifies the boundary between the two candidate growth regimes. In an unbalanced-growth equilibrium, Lemma \ref{lem:G>1} shows that the interest rate equals the economic growth rate, $R=G>1$. Because $\bar z=R/m$, equation \eqref{eq:return_growth_gap} and the uniqueness of the root in Lemma \ref{lem:leverage_threshold} imply that $\bar z=v_\lambda$ and $G=mv_\lambda$. The boundary candidate therefore has $v_{\bar\lambda}=1/m$ and $G=R=1$. For $\lambda>\bar\lambda$, the associated candidate growth rate satisfies $G=mv_\lambda>1$.

Although $(\bar\lambda,1/m)$ is admissible, for $\lambda>\bar\lambda$ the root $v_\lambda$ defines a valid equilibrium cutoff only if
\begin{equation*}
	\lambda\Phi(v_\lambda)+1-\lambda>0.
\end{equation*}
By continuity, Assumption \ref{asmp:bmz} guarantees this condition for $\lambda$ in a right neighborhood of $\bar\lambda$. \S\ref{sec:loose_constraints} characterizes the restriction for arbitrary leverage. The following theorem characterizes both growth regimes.

\begin{thm}[Growth regimes and land bubbles: benchmark model]\label{thm:trend}
Suppose Assumptions \ref{asmp:Phi}, \ref{asmp:bmz} hold, let $\bar{\lambda}$ be as in \eqref{eq:lambda_threshold}, and fix $K_0>0$. Then:
\begin{enumerate}
    \item\label{item:trend_f} A balanced-growth equilibrium exists if and only if $\lambda<\bar{\lambda}$ and, for the unique $R\in(1,1/\beta)$ satisfying $1=\beta(R+\lambda\pi(R))$,
    \begin{equation}
        K_0=\frac{\lambda D}{(R-1)[\lambda\Phi(R/m)+1-\lambda]}
        \int_{R/m}^\infty z\diff\Phi(z). \label{eq:K0_balanced}
    \end{equation}
    When it exists, it is unique and satisfies $G=1$, $N_t=0$, $P_t=D/(R-1)$, and hence is fundamental.
    \item\label{item:trend_b} An unbalanced-growth equilibrium exists if and only if $\lambda>\bar{\lambda}$ and $v_\lambda$ is admissible. When it exists, it is unique and satisfies $\bar z=v_\lambda$, $G=R=mv_\lambda>1$, $N_t=-D/(R-1)$, and, writing $b\coloneqq\beta(\lambda\Phi(\bar z)+1-\lambda)$,
    \begin{equation}
        P_t=\frac{b}{1-b}R^t\left(mK_0+\frac{RD}{R-1}\right)+\frac{D}{R-1}. \label{eq:Pt_b}
    \end{equation}
    This equilibrium is bubbly.
    \item If $\lambda=\bar\lambda$, neither a balanced-growth nor an unbalanced-growth equilibrium as defined above exists.
\end{enumerate}
\end{thm}

The conclusion $N_t=-D/(R-1)$ in part \ref{item:trend_b} does not conflict with the zero initial position $N_{-1}=0$. Agents choose their date-0 bond positions after entering with their initial portfolios, and in the unbalanced-growth equilibrium their aggregate borrowing from the financial sector equals the fundamental value $D/(R-1)$. This borrowing finances the fundamental component of land value, while the bubble component supplies the remaining net demand for safe assets.

The boundary candidate within the constant-growth, constant-interest-rate formulation has $G=R=1$.\footnote{In the related simplified model analyzed by \citet[\S6, Proposition~7(ii)]{HiranoToda2024JME}, the corresponding knife-edge case admits a subexponential equilibrium path: land rent is constant, the land price---and hence the price--rent ratio---grows linearly, and $P_t=V_t$ at every date, so there is no bubble. This is consistent with Theorem \ref{thm:trend}: that path belongs to neither equilibrium class defined above and does not feature the constant $R>1$ maintained in Definition \ref{defn:lr}.} We deliberately exclude this knife-edge case because a constant positive dividend has no finite fundamental value under a constant $R=1$; the linear benchmark therefore has no equilibrium in either class at the threshold. \S\ref{sec:general} instead has endogenous rents and studies the two economically relevant regimes on either side of the threshold.

\section{Financial conditions and bubble necessity}\label{sec:financial_conditions}

In this section, we separate two questions that are often conflated. Bubble possibility asks whether at least one bubbly equilibrium exists under given financial conditions. Bubble necessity is stronger: an equilibrium exists and every equilibrium is bubbly, so fundamental valuation is impossible. In our model, these questions require two steps. We first characterize when the post-threshold unbalanced-growth equilibrium exists. We then show that whenever it exists, it is unique in the class considered in Theorem~\ref{thm:trend} and land is necessarily bubbly because no fundamental equilibrium exists.

\subsection{Why relaxing leverage need not eliminate bubbles}\label{sec:loose_constraints}

We begin by establishing when the post-threshold unbalanced-growth equilibrium exists. This step is logically prior to bubble necessity: if the candidate is inadmissible and no equilibrium exists, a statement that every equilibrium is bubbly would be vacuous. Theorem \ref{thm:trend} reduces existence to admissibility of the root $v_\lambda$ characterized in Lemma \ref{lem:leverage_threshold}. We now characterize that restriction using the conditional tail expectation. Assumption \ref{asmp:bmz} guarantees that the bubble necessity region begins when leverage crosses $\bar\lambda$, but the productivity distribution determines how far that region extends. This characterization also explains why our conclusion differs from the usual result in the literature that sufficiently developed financial markets eliminate bubbly equilibria.

\paragraph{Conditional tail expectation}
Let $z_{\max}\in(0,\infty]$ be the upper endpoint of the support of $\Phi$. For any $v$ such that $1-\Phi(v)>0$, we refer to $\E[Z\mid Z\ge v]$ as the \emph{conditional tail expectation} at $v$. The defining condition for this root can be written as
\begin{equation}
    (1-\beta)v_\lambda
    =\beta\lambda[1-\Phi(v_\lambda)]
    \left(\E[Z\mid Z\ge v_\lambda]-v_\lambda\right).
    \label{eq:root_conditional_mean}
\end{equation}
Because admissibility is equivalent to $\lambda[1-\Phi(v_\lambda)]<1$, equation \eqref{eq:root_conditional_mean} immediately yields the following result.

\begin{prop}[Existence of unbalanced growth]\label{prop:admissible_leverage}
For $\lambda>\bar\lambda$, let $v_\lambda>1/m$ be the root characterized in Lemma \ref{lem:leverage_threshold}. Then:
\begin{enumerate}
    \item\label{item:admissible_iff} The candidate cutoff is admissible if and only if
    \begin{equation}
        \beta\E[Z\mid Z\ge v_\lambda]>v_\lambda.
        \label{eq:conditional_mean_criterion}
    \end{equation}
    \item\label{item:v_to_zmax} As $\lambda\to\infty$, we have $v_\lambda\to z_{\max}$.
    \item\label{item:thin_tail} If
    \begin{equation}
        \limsup_{v\to z_{\max}}
        \frac{\E[Z\mid Z\ge v]}{v}<\frac{1}{\beta},
        \label{eq:thin_tail_condition}
    \end{equation}
    then the candidate is inadmissible for all sufficiently large $\lambda$. In particular, \eqref{eq:thin_tail_condition} holds if $Z$ has bounded support.
    \item\label{item:thick_tail} If
    \begin{equation}
        \liminf_{v\to z_{\max}}
        \frac{\E[Z\mid Z\ge v]}{v}>\frac{1}{\beta},
        \label{eq:thick_tail_condition}
    \end{equation}
    then the candidate is admissible for all sufficiently large $\lambda$. If $\beta\E[Z\mid Z\ge v]>v$ for every $v\ge1/m$ in the support, it is admissible for every $\lambda>\bar\lambda$.
\end{enumerate}
\end{prop}

\begin{proof}
At the root, substituting \eqref{eq:root_conditional_mean} into $\lambda[1-\Phi(v_\lambda)]<1$ and rearranging gives \eqref{eq:conditional_mean_criterion}, proving part \ref{item:admissible_iff}. By Lemma \ref{lem:leverage_threshold}, $v_\lambda$ is increasing in leverage. If it converged to some $v<z_{\max}$ as $\lambda\to\infty$, then the integral in \eqref{eq:Psi} would remain positive, contradicting the root condition $\Psi(v_\lambda,\lambda)=0$. This proves part \ref{item:v_to_zmax}. Parts \ref{item:thin_tail} and \ref{item:thick_tail} now follow directly from parts \ref{item:admissible_iff} and \ref{item:v_to_zmax}. If $z_{\max}<\infty$, then $1\le\E[Z\mid Z\ge v]/v\le z_{\max}/v\to 1$ as $v\to z_{\max}$. Because $1<1/\beta$, condition \eqref{eq:thin_tail_condition} holds.
\end{proof}

Assumption \ref{asmp:bmz} is exactly criterion \eqref{eq:conditional_mean_criterion} at $v_{\bar\lambda}=1/m$. By continuity, it guarantees that the unbalanced-growth region contains a nonempty interval immediately above $\bar\lambda$. It does not assert that the candidate remains admissible at arbitrarily large leverage. The exact unbalanced-growth region, which is also the bubble necessity region, is
\begin{equation}
    \cL_B
    =\set*{\lambda>\bar\lambda:
    \beta\E[Z\mid Z\ge v_\lambda]>v_\lambda}.
    \label{eq:bubbly_leverage_set}
\end{equation}
For every $\lambda\in\cL_B$, Theorem \ref{thm:trend} implies that the unique equilibrium in the class considered here exhibits unbalanced growth and a land bubble. Thus $\cL_B$ is not merely a parameter region in which one bubbly equilibrium is possible: no fundamental equilibrium in this class coexists with the bubbly equilibrium.

\paragraph{Bounded support and the two-type benchmark}
Proposition \ref{prop:admissible_leverage}\ref{item:thin_tail} shows that bounded productivity eventually reproduces the standard result in the literature that sufficiently high financial development rules out bubbles. As leverage becomes large, the investment cutoff approaches the most productive technology and the conditional tail expectation becomes arbitrarily close to the cutoff. Because agents consume the fraction $1-\beta$ of wealth, growth then falls below the return required by savers, and the equilibrium with unbalanced growth ceases to exist.

This is the force operating in the two-type model of \citet[Proposition~1 and eq.~(15)]{HiranoYanagawa2017}: for $\theta\ge1-p$, the safe return is $\alpha^H$ while growth is $g=\beta\alpha^H$. Because $\beta<1$, their necessary and sufficient bubble condition $g\ge r$ \citep[Proposition~3]{HiranoYanagawa2017} fails. Their continuum extension assumes productivity is bounded between $\ubar{\alpha}$ and $\bar\alpha$; its market-clearing and growth equations \citep[Appendix~A.12, eqs.~(34)--(35)]{HiranoYanagawa2017} are used to obtain the analogous conclusion at high pledgeability. Thus the contrast is not simply ``two types versus a continuum.'' The relevant distinction is whether the conditional tail expectation remains sufficiently far above an increasingly high cutoff.

The borrowing constraints also differ. \citet{HiranoYanagawa2017} restrict debt repayment to a fraction of project output, so effective leverage depends on productivity and the endogenous interest rate. Our constraint directly limits investment relative to equity. This changes the equilibrium formulas, but Proposition \ref{prop:admissible_leverage} shows that the constraint specification alone does not overturn the bounded-support conclusion.

\paragraph{Exponential and Pareto productivity}
We illustrate the role of upper-tail behavior using two distributions that are especially relevant for our comparison. Suppose first that productivity is exponentially distributed with decay rate $\gamma>0$:
\begin{equation}
    \Phi(z)=1-\e^{-\gamma z}. \label{eq:exp}
\end{equation}
Its conditional tail expectation is $\E[Z\mid Z\ge v]=v+1/\gamma$. By Proposition \ref{prop:admissible_leverage}\ref{item:admissible_iff}, the candidate is admissible if and only if $v<\beta/[\gamma(1-\beta)]$. Combining this condition with $\Psi(v,\lambda)=0$ gives
\begin{equation}
    \bar\lambda<\lambda<\lambda_{\max}\coloneqq \exp\left(\frac{\beta}{1-\beta}\right).
    \label{eq:exponential_lambda_max}
\end{equation}
Hence exponential productivity does not support the unbalanced-growth equilibrium for literally arbitrary leverage. Quantitatively, however, the upper bound is enormous: even with a small discount factor like $\beta=0.9$, we have $\lambda_{\max}=\exp(9)\approx 8.1\times10^3$, far above realistic values.

By contrast, suppose productivity has a Pareto upper tail with exponent $\zeta>1$. For cutoffs in that tail, its conditional tail expectation is
\begin{equation*}
    \E[Z\mid Z\ge v]=\frac{\zeta}{\zeta-1}v.
\end{equation*}
By Proposition \ref{prop:admissible_leverage}\ref{item:admissible_iff}, $v_\lambda$ is admissible at every such cutoff if and only if
\begin{equation}
    \beta\frac{\zeta}{\zeta-1}>1 \iff \zeta<\frac{1}{1-\beta}.
    \label{eq:Pareto_admissibility}
\end{equation}
In this case the bubble necessity region is unbounded above. What matters is not unbounded support by itself, but that the conditional tail expectation remains a constant proportion above the cutoff.

The Pareto case connects directly to \citet{BiswasHansonPhan2020}. Identifying their net borrowing $d$ with $-a_{it}$ and their investment $I$ with $i_{it}$, their borrowing constraint $d\le\theta I$ is equivalent to \eqref{eq:borrowing_constraint} with $\theta=1-1/\lambda$. Under Pareto productivity, equations (20), (24), and (25) of \citet{BiswasHansonPhan2020} imply that their bubbleless interest rate is $(\zeta-1)/(\beta\zeta)$, independent of $\theta$; their Lemma~2 then shows that, when the pure bubble survives with probability one, the bubble existence condition is exactly \eqref{eq:Pareto_admissibility}. Their model therefore contains the same upper-tail force: relaxing the constraint moves the cutoff farther into a scale-invariant Pareto tail without eliminating demand for the bubbly asset.

\paragraph{Binding constraints and constraint tightness}
The result separates the existence of a binding financial constraint from the tightness of that constraint. Whenever an unbalanced-growth equilibrium exists, high-productivity agents invest up to the constraint and, when $\lambda>1$, borrow up to the limit, while low-productivity agents demand safe claims. Relaxing a constraint that remains binding can nevertheless increase investment and growth enough to move the economy through $\bar\lambda$. The conclusion is therefore not that bubbles require unconstrained finance, but that making a constraint tighter need not make bubbles more likely.

Proposition \ref{prop:admissible_leverage} answers the existence question for the post-threshold regime: it determines when the candidate unbalanced-growth allocation defines an equilibrium. It does not by itself establish bubble necessity. Theorem \ref{thm:trend} and the next subsection do so by combining the financial mechanism with the positive dividend and the change from balanced to unbalanced growth. Thus the productivity tail determines whether the bubble necessity regime exists, not whether a bubble is merely one possible equilibrium within that regime.

\subsection{Phase transition, interest rates, and bubble necessity}\label{sec:bubble_necessity}

The preceding subsection identifies when an unbalanced-growth equilibrium exists. We now explain why land is necessarily bubbly in that equilibrium. The claim is not that leverage makes a bubble supportable among several equilibria. Rather, once $\lambda>\bar\lambda$ and the unbalanced-growth equilibrium is admissible, the fundamental balanced-growth equilibrium is impossible and the unique equilibrium in the class considered in Theorem~\ref{thm:trend} is bubbly. The mechanism has three steps. First, low-productivity agents save in safe claims, while high-productivity agents invest up to their leverage limits and finance part of that investment by issuing safe claims. At finite leverage, this intermediation leaves positive aggregate net demand for safe claims. Second, land is the only safe asset in fixed positive supply, while the financial sector's long-run market-clearing requirement prevents it from carrying a position of the same order as aggregate wealth. When leverage rises in the balanced-growth regime, the land price increases and the interest rate falls. Third, beyond $\bar\lambda$, the balanced-growth equation would require a gross interest rate at or below one, which is incompatible with a finite fundamental value of a positive constant rent. The price of land must then grow with the economy to provide the additional store of value.

The requirement $N_t/W_t\to0$, discussed after Definition \ref{defn:lr}, determines the constant interest rate. \S\ref{sec:general} establishes analogous stationary and locally unbalanced regimes under exact clearing among agents, showing that the mechanism does not depend on the benchmark financial sector.

The following proposition characterizes the interest-rate adjustment underlying this transition.

\begin{prop}[V-shaped interest rate]\label{prop:rate}
Let everything be as in Theorem \ref{thm:trend}. The equilibrium gross risk-free rate $R$ is strictly decreasing in $\lambda$ in the balanced-growth region and strictly increasing in $\lambda$ in the unbalanced-growth region. Moreover, $R\to1$ as $\lambda\to\bar\lambda$ from either side.
\end{prop}

\begin{figure}[!htb]
\centering
\includegraphics[width=0.7\linewidth]{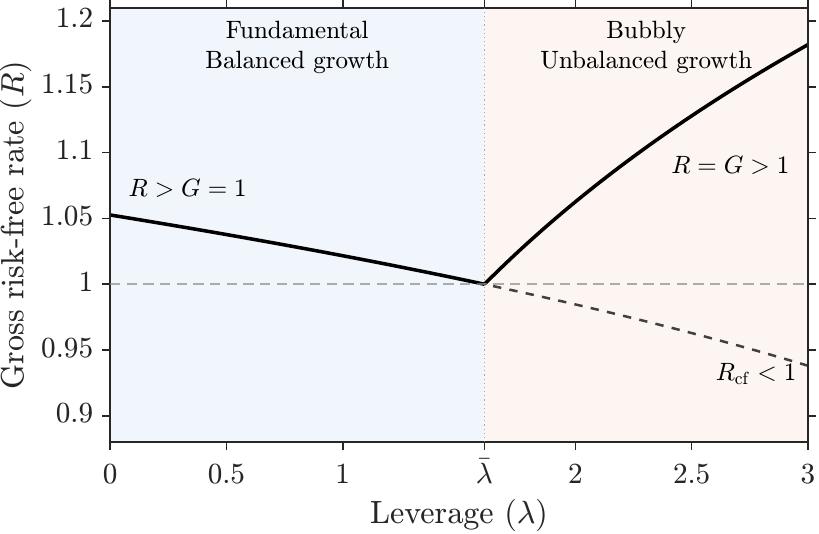}
\caption{Interest rates in balanced-growth and unbalanced-growth equilibria. The dashed curve shows the counterfactual interest rate implied by the balanced-growth equation for $\lambda>\bar\lambda$. Parameter values are specified in \S\ref{sec:discuss}.}\label{fig:rate}
\end{figure}

Proposition \ref{prop:rate} shows that the relation between leverage and interest rates changes at the phase transition. In the balanced-growth region, the equilibrium condition fixes $G=1$. Holding $R$ fixed, relaxing leverage allows productive agents to borrow more and expand investment, increasing the contribution $\lambda\pi(R)$ in \eqref{eq:G} and thereby putting upward pressure on aggregate wealth growth. Because balanced growth requires $G$ to remain equal to one, equilibrium offsets this force by lowering $R$, the return on safe claims earned by unproductive agents. Thus relaxing leverage lowers the interest rate without affecting long-run economic growth. As $\lambda\uparrow\bar\lambda$, the gross interest rate falls toward one. In the unbalanced-growth region, by contrast, $G=R=mv_\lambda>1$ and $v_\lambda$ increases with leverage. Thus further relaxation raises both the economic growth rate and the rate at which the bubble component grows. Low interest rates characterize the approach to the transition from the balanced-growth side, not the entire bubbly regime. Figure~\ref{fig:rate} illustrates the V-shaped interest rate and the associated phase transition.

\citet{HiranoToda2025JPE} proposed the concept of the \emph{necessity} of asset price bubbles and proved that bubbles necessarily arise in certain workhorse macroeconomic models whenever the bubble necessity condition
\begin{equation}
    R_\mathrm{cf}<G_d<G \label{eq:necessity}
\end{equation}
    holds, where $G$ is the economic growth rate, $G_d$ is the dividend growth rate, and $R_\mathrm{cf}$ is the counterfactual bubbleless interest rate. Figure~\ref{fig:rate} makes this condition transparent. Constant rents imply $G_d=1$. As leverage approaches $\bar\lambda$ from below, the equilibrium interest rate falls toward one. Continuing the balanced-growth equation beyond the threshold yields the dashed curve $R_\mathrm{cf}<1$, while the actual unbalanced-growth equilibrium has $G=R>1$. Hence the bubble necessity condition \eqref{eq:necessity} holds. Moreover, because $D>0$ is constant, the fundamental value of the asset would be infinite under $R_\mathrm{cf}\le1$. A fundamental equilibrium therefore cannot be sustained beyond the threshold; any equilibrium in the class considered in Theorem~\ref{thm:trend} must exhibit unbalanced growth and an asset price bubble.

The role of $D>0$ in Theorem \ref{thm:trend} is to discipline land valuation, not to generate the leverage threshold. A positive rent makes land a dividend-paying asset and pins down the strictly positive fundamental component $V=D/(R-1)$ when $R>1$. The theorem can therefore establish bubble necessity for an asset with positive fundamental value. Accordingly, $D$ does not affect the leverage threshold \eqref{eq:lambda_threshold} or the growth mechanism; in the unbalanced-growth equilibrium, it affects the level of the fundamental component but not asymptotic growth.

The phase transition also changes the role of financial conditions in determining economic growth. Below $\bar\lambda$, the economy is on a balanced-growth path with $G=1$: leverage changes the interest rate, land price, and allocation of investment, but not the long-run growth rate. In the unbalanced-growth region, leverage determines $v_\lambda$ and therefore the endogenous growth rate $G=mv_\lambda>1$. The phase transition is consequently not only a switch from fundamental to bubbly valuation; it is also a switch in the economy's long-run growth behavior.

Using the explicit expression for the leverage threshold \eqref{eq:lambda_threshold} separating the two growth regimes, we obtain the following comparative statics. 

\begin{prop}[Comparative statics of leverage threshold]\label{prop:compstat}
Let $\bar{\lambda}(\beta,m,\Phi)$ be the leverage threshold in \eqref{eq:lambda_threshold}. Then $\bar{\lambda}$ is strictly decreasing in $\beta$ and $m$. It is decreasing under upward shifts in productivity in the sense of first-order stochastic dominance: if $\Phi_1 \preceq \Phi_2$, meaning that $1-\Phi_1(z)\le 1-\Phi_2(z)$ for all $z$, then $\bar{\lambda}(\beta,m,\Phi_2)\le\bar{\lambda}(\beta,m,\Phi_1)$. 
\end{prop}

\begin{proof}
The strict monotonicity of $\bar{\lambda}$ in $\beta$ and $m$ follows from \eqref{eq:lambda_threshold}. The monotonicity with respect to $\Phi$ follows from the monotonicity of $z\mapsto \max\set{mz-1,0}$ and the definition of first-order stochastic dominance.
\end{proof}

Proposition \ref{prop:compstat} implies that greater patience and a higher asymptotic marginal product of capital decrease the leverage threshold separating balanced from unbalanced growth, while higher productivity weakly decreases it. Intuitively, these changes strengthen the positive feedback loop between capital investment and land prices, so less leverage is required for economic takeoff.

The two regimes differ in what anchors the land price. In the balanced-growth equilibrium, $P_t=V=D/(R-1)$: the price adjusts to equal the present discounted value of rents. We call this valuation \emph{supply-driven} because the asset's cash flow fully anchors its price. The rent $D$ is fixed in the tractable benchmark and is determined by the production technology in \S\ref{sec:general}. Because the interest rate is endogenous, asset demand still affects the discount rate and hence the level of $V$; ``supply-driven'' means that no component of the price is left unsupported by rents. In the unbalanced-growth equilibrium, by contrast, land valuation becomes \emph{demand-driven}. Ex ante heterogeneity generates in every period a group of low-productivity agents who do not invest in capital and instead save in safe claims. Equation \eqref{eq:Pt1} shows that aggregate net demand for safe claims is the positive fraction $b\coloneqq\beta(\lambda\Phi(\bar z)+1-\lambda)$ of aggregate wealth. After productive agents finance investment by issuing safe claims, the economy therefore retains positive aggregate net demand for them, and land is the only safe asset in positive net supply. Although the financial sector can accommodate agents' aggregate net bond position at any finite date, the long-run market-clearing requirement $N_t/W_t\to0$ prevents the sector's opposite position from remaining a nonvanishing share of wealth. Because land is in unit supply, market clearing therefore requires $P_t/W_t\to b$: the land price adjusts to the demand for a store of value and grows with aggregate wealth. Correspondingly, \eqref{eq:Pt_b} decomposes $P_t$ into the constant fundamental value $V$ and a component proportional to $R^t$. Since rents remain constant, this wealth-driven component is a bubble.

This distinction generates an observable difference in price--dividend dynamics. In balanced growth, the price--dividend ratio $P_t/D$ is constant. In unbalanced growth, the dividend is constant while $P_t$ grows at rate $G>1$, so $P_t/D$ grows without bound. The model thereby provides a structural rationale for the explosive price--dividend behavior used to detect bubbles empirically \citep{PhillipsShiYu2015}. More broadly, \citet[p.~1]{Barlevy2018} discusses why rapid asset-price increases unaccompanied by corresponding changes in expected dividends concern policymakers.

\section{General model}\label{sec:general}

\S\ref{sec:simple} isolates the mechanism in a tractable benchmark. This section introduces a general production technology with endogenous land rents and exact market clearing among agents. The extension serves two purposes. First, it shows that the same leverage threshold separates a unique steady state at low leverage from the absence of stationary equilibria at high leverage and, under additional conditions, constructs a locally unique unbalanced-growth equilibrium on the high-leverage side. Second, it identifies the condition under which endogenous rents grow too slowly to support the land price in that unbalanced-growth regime.

\subsection{Model}

The model is identical to \S\ref{sec:simple} except that
\begin{enumerate*}
    \item capital and land are combined through a general production technology,
    \item land rents are endogenous, and
    \item there is no unmodeled financial sector, so all positions in risk-free claims clear among agents.
\end{enumerate*}

\paragraph{Production}
Production uses capital (denoted by $K$) and land (denoted by $X$). There is a representative firm with a neoclassical constant-returns-to-scale production function $Y(K,X)$. Factor markets are competitive, and inputs are paid their marginal products. After production, capital depreciates at rate $\delta\in [0,1)$. To simplify notation, following \citet{Coleman1991}, we introduce the function
\begin{equation*}
    F(K,X)\coloneqq Y(K,X)+(1-\delta)K.
\end{equation*}
The equilibrium can be described entirely in terms of $F$, which is the non-land component of aggregate wealth. For any differentiable function $F$ that is homogeneous of degree 1 and has positive partial derivatives, Euler's theorem implies
\begin{equation*}
    \frac{\diff}{\diff K}\frac{F(K,1)}{K}=-\frac{F_X(K,1)}{K^2}<0.
\end{equation*}
Thus the average product $F(K,1)/K$ is strictly decreasing. We impose the following regularity and boundary conditions.

\begin{asmp}\label{asmp:F}
$F:\R_{++}^2\to \R_+$ is homogeneous of degree 1, concave, twice continuously differentiable with positive partial derivatives, and satisfies
\begin{subequations}
\begin{align}
    \lim_{K\to\infty}\frac{F(K,1)}{K}
    &=\lim_{K\to\infty}F_K(K,1)\eqqcolon m>0, \label{eq:MPK}\\
    \lim_{K\downarrow 0}\frac{F(K,1)}{K}
    &\eqqcolon \bar{m}\in(m,\infty]. \label{eq:mbar}
\end{align}
\end{subequations}
\end{asmp}
The small-capital limit in \eqref{eq:mbar} exists by the preceding monotonicity. When $\bar{m}<\infty$, concavity additionally implies $\lim_{K\downarrow0}F_K(K,1)=\bar{m}$; we use this implication in the proof of Proposition \ref{prop:asymptotic}.

The tractable model in \S\ref{sec:simple} corresponds to the special case $F(K,X)=mK+DX$, for which $\bar{m}=\infty$. The positivity of $m$ follows from $F_K=Y_K+1-\delta>1-\delta>0$, which implies $m\ge 1-\delta>0$. The asymptotic slope $m$ is the familiar asymptotic-$AK$ channel for endogenous growth.\footnote{For the CES production function \eqref{eq:CES} with elasticity of substitution $1/\rho>1$, the average and marginal products of capital converge to a positive constant as $K\to\infty$. This possibility for a general production function was recognized by \citet[p.~72]{Solow1956}; see also \citet[pp.~68--69]{BarroSala-i-Martin2004}. \citet{JonesManuelli1990} study endogenous growth along this line, with their condition G playing a role analogous to Assumption \ref{asmp:bmz}. Here $m$ is positive for every value of $\rho$ because $F$ includes undepreciated capital. Our contribution is to connect this growth channel to leverage and to the transition between asymptotically balanced and unbalanced growth.} A typical example satisfying Assumption \ref{asmp:F} is the constant elasticity of substitution (CES) production function
\begin{equation}
    Y(K,X)=\begin{cases*}
        A\left(\alpha K^{1-\rho}+(1-\alpha)X^{1-\rho}\right)^\frac{1}{1-\rho} & if $0<\rho\neq 1$,\\
        AK^\alpha X^{1-\alpha} & if $\rho=1$,
    \end{cases*}\label{eq:CES}
\end{equation}
where $A>0$, $\alpha\in(0,1)$ are parameters and $\rho>0$ is the reciprocal of the elasticity of substitution between capital and land. In this case, a straightforward calculation yields
\begin{equation*}
	 m=\begin{cases*}
	 A\alpha^\frac{1}{1-\rho}+1-\delta & if $\rho<1$,\\
	 1-\delta & if $\rho\ge 1$.
	 \end{cases*}
\end{equation*}
The corresponding small-capital limit is
\begin{equation}
    \bar{m}=\begin{cases*}
        \infty & if $\rho\le 1$,\\
        A\alpha^\frac{1}{1-\rho}+1-\delta & if $\rho>1$.
    \end{cases*} \label{eq:mbar_CES}
\end{equation}

\paragraph{Land}

The aggregate supply of land is exogenous and normalized to 1. Agents can take arbitrary positive or negative positions in claims to land. Because we restrict attention to deterministic aggregate equilibria, the return on these claims is risk-free. Appendix \ref{sec:land_trading} describes an equivalent implementation using a real estate investment trust (REIT).

The gross risk-free rate between time $t$ and $t+1$, denoted by $R_t$, is determined as follows. Let $K_t$ and $P_t$ denote aggregate capital and the land price (excluding current rent) at time $t$. Because the aggregate supply of land is 1, the aggregate land rent at time $t+1$ is $F_X(K_{t+1},1)$. Therefore, the gross risk-free rate (return on land) is
\begin{equation}
    R_t\coloneqq \frac{P_{t+1}+F_X(K_{t+1},1)}{P_t}. \label{eq:Rt}
\end{equation}

\paragraph{Individual behavior}

Individual behavior is the same as in \S\ref{sec:simple}, except that the interest rate $R_t$ is time-varying as in \eqref{eq:Rt} and the reduced law of motion for wealth in \eqref{eq:budget_t+_reduced} becomes
\begin{equation}
    w_{t+1}\coloneqq \underbrace{F_K(K_{t+1},1)z_t i_t}_\text{income from capital}+\underbrace{R_ta_t}_\text{income from asset}. \label{eq:budget_t+1}
\end{equation}
Consequently, the productivity threshold for investment is determined as
\begin{equation}
    F_K(K_{t+1},1)z_t>R_t\iff z_t>\bar{z}_t\coloneqq \frac{R_t}{F_K(K_{t+1},1)}. \label{eq:z_threshold}
\end{equation}

\subsection{Equilibrium dynamics and financial accelerator}

\paragraph{Equilibrium conditions and dynamics}

Because the aggregate supply of land is 1, the market value of outstanding land claims equals the land price $P_t$. Aggregating date-$t$ wealth (\eqref{eq:budget_t+1} with time shifted by 1) across agents and using the definition of the risk-free rate \eqref{eq:Rt}, we obtain
\begin{align}
    W_t&\coloneqq F_K(K_t,1)K_t+R_{t-1}P_{t-1} \notag \\
    &=F_K(K_t,1)K_t+P_t+F_X(K_t,1)=F(K_t,1)+P_t,\label{eq:Wt}
\end{align}
where the last equality uses the homogeneity of $F$ (Assumption \ref{asmp:F}). The first equality applies for $t\ge1$; at $t=0$, the applicable accounting identity is $W_0=F(K_0,1)+P_0$. Thus $F$, rather than $Y$, is the non-land component of aggregate wealth. Equations \eqref{eq:Kt+} and \eqref{eq:Pt1} remain valid after setting $\bar{z}=\bar{z}_t$ as in \eqref{eq:z_threshold} and $N_t=0$, reflecting the absence of positions with an unmodeled financial sector. We may therefore define a rational expectations equilibrium as follows.

\begin{defn}\label{defn:REE}
Given the initial aggregate capital $K_0>0$, a \emph{rational expectations equilibrium} consists of a sequence $\set{(R_t,\bar{z}_t,W_t,K_{t+1},P_t)}_{t=0}^\infty$ of gross risk-free rates, productivity thresholds, aggregate wealth levels, aggregate capital stocks, and land prices such that
\begin{enumerate*}
    \item $R_t>0$ satisfies \eqref{eq:Rt},
    \item $\bar{z}_t>0$ is admissible and satisfies \eqref{eq:z_threshold},
    \item $W_t>0$ satisfies \eqref{eq:Wt},
    \item $K_{t+1}>0$ satisfies \eqref{eq:Kt+} with $\bar{z}=\bar{z}_t$, and
    \item $P_t>0$ satisfies \eqref{eq:Pt1} with $N_t=0$.
\end{enumerate*}
\end{defn}

Definition \ref{defn:REE} gives five nonlinear difference equations in five unknowns. The following lemma reduces them to a two-dimensional system.

\begin{lem}[Two-dimensional equilibrium dynamics]\label{lem:dynamics}
Suppose Assumptions \ref{asmp:Phi}, \ref{asmp:F} hold and define the functions
\begin{subequations}\label{eq:WPKz}
\begin{align}
    W(K,v)&\coloneqq \frac{1}{1-\beta(\lambda\Phi(v)+1-\lambda)}F(K,1), \label{eq:WKz}\\
    P(K,v)&\coloneqq \frac{\beta(\lambda\Phi(v)+1-\lambda)}{1-\beta(\lambda\Phi(v)+1-\lambda)}F(K,1), \label{eq:PKz}
\end{align}
\end{subequations}
where we restrict the domain to $K>0$ and admissible $v$ so that $W,P>0$. Given the initial aggregate capital $K_0$, the equilibrium is characterized by the two-dimensional system
\begin{subequations}\label{eq:dynamics}
    \begin{align}
        K_{t+1}&=\beta\lambda W(K_t,\bar{z}_t)\int_{\bar{z}_t}^\infty z\diff \Phi(z), \label{eq:dynamics_K} \\
        \bar{z}_t&=\frac{P(K_{t+1},\bar{z}_{t+1})+F_X(K_{t+1},1)}{F_K(K_{t+1},1)P(K_t,\bar{z}_t)}. \label{eq:dynamics_z}
    \end{align}
\end{subequations}
\end{lem}

\begin{proof}
Using \eqref{eq:Wt} and \eqref{eq:Pt1} with $N_t=0$, we may solve for $W_t$ and $P_t$, which yields $W_t=W(K_t,\bar{z}_t)$ and $P_t=P(K_t,\bar{z}_t)$ in \eqref{eq:WPKz}. Substituting these equations into \eqref{eq:Kt+}, we obtain \eqref{eq:dynamics_K}. Finally, \eqref{eq:dynamics_z} follows from the definitions of $R_t$ in \eqref{eq:Rt} and $\bar{z}_t$ in \eqref{eq:z_threshold}.
\end{proof}

\paragraph{Financial accelerator}

The equilibrium equations reveal the same positive feedback loop between the land price and the real economy as in \S\ref{sec:simple}, now with an endogenous land rent. Agents with low realized productivity prefer the risk-free land claim, whereas high-productivity agents borrow against their net worth to invest in capital. Because land is the only risk-free asset in positive net supply, its fixed supply converts the resulting demand for safe saving into a higher price. A higher land price raises current aggregate wealth one-for-one by \eqref{eq:Wt}. For a given productivity threshold, the increase in wealth finances more capital investment by \eqref{eq:Kt+}, with the response amplified by leverage $\lambda$. Greater capital in turn raises production and weakly raises the land rent, thereby supporting future wealth. Finally, by \eqref{eq:Pt1} with $N_t=0$, a fraction of aggregate wealth is allocated to land, supporting its price and closing the feedback loop. The asymptotic analysis below determines whether this feedback subsides as the economy converges to a steady state or is strong enough to sustain an unbalanced-growth regime. In the latter case, the central asset-pricing question is whether the induced growth in land rents is sufficient to support the growth in land prices.

\paragraph{Small-capital productivity}

We now impose an additional condition that guarantees that the average product becomes sufficiently high when capital is scarce. To motivate the condition, consider a steady state with capital $K>0$ and productivity threshold $v$. Setting $K_{t+1}=K_t=K$ and $\bar{z}_t=v$ in \eqref{eq:dynamics_K} and substituting \eqref{eq:WKz}, we obtain
\begin{equation}
    \frac{F(K,1)}{K}
    =\frac{1-\beta(\lambda\Phi(v)+1-\lambda)}{\beta\lambda\int_v^\infty z\diff\Phi(z)}. \label{eq:required_AP}
\end{equation}
At the reference threshold $v=1/m$, the average product required by \eqref{eq:required_AP} is
\begin{equation*}
    \frac{1-\beta(\lambda\Phi(1/m)+1-\lambda)}{\beta\lambda\int_{1/m}^\infty z\diff\Phi(z)}
    =\frac{1-\beta}{\beta\lambda\int_{1/m}^\infty z\diff\Phi(z)}
    +\frac{1-\Phi(1/m)}{\int_{1/m}^\infty z\diff\Phi(z)}.
\end{equation*}
This expression is strictly decreasing in $\lambda\ge 1$ and hence is largest when $\lambda=1$. We therefore impose the following assumption.

\begin{asmp}\label{asmp:smallK}
The model parameters satisfy
\begin{equation*}
    \beta\left(\Phi(1/m)+\bar{m}\int_{1/m}^\infty z\diff\Phi(z)\right)>1.
\end{equation*}
\end{asmp}

Assumption \ref{asmp:smallK} is equivalent to requiring $\bar{m}$ to exceed the value on the right-hand side of \eqref{eq:required_AP} at $v=1/m$ and $\lambda=1$. Thus, even under the tightest leverage constraint, capital becomes productive enough when scarce for the stationary capital equation to have a positive solution around the reference threshold. Because the required average product is largest at $\lambda=1$, the assumption is uniform over leverage. Assumptions \ref{asmp:bmz} and \ref{asmp:smallK} play complementary roles: the former ensures that sufficiently productive agents can generate growth at the asymptotic productivity $m$, whereas the latter ensures that scarce capital has a sufficiently high average product to support a stationary equilibrium. Assumption \ref{asmp:smallK} holds automatically in the linear model because $\bar{m}=\infty$. By \eqref{eq:mbar_CES}, it also holds automatically for the CES production function when $\rho\le 1$ and imposes a finite lower bound on the small-capital average product when $\rho>1$.

\subsection{Asymptotic behavior}

We focus on two economically relevant classes of long-run equilibria. In an \emph{asymptotically balanced-growth equilibrium}, the economy converges to an admissible steady state and land prices and rents have the same asymptotic growth rate. The qualifier ``asymptotically'' reflects that the general model typically undergoes transition dynamics before converging to the steady state; in the tractable benchmark of \S\ref{sec:simple}, the corresponding equilibrium lies on its balanced-growth path from the outset. In an \emph{unbalanced-growth equilibrium}, capital and land prices exhibit sustained endogenous growth while land rents grow more slowly. We study how leverage affects the existence and asset-pricing properties of these two regimes; we do not claim a global classification of all equilibrium paths. Because our model features homothetic preferences, a constant-returns-to-scale production function, and a production factor in fixed supply (land), one might expect land eventually to constrain growth. We show that this intuition is incorrect and an unbalanced-growth regime can instead arise despite the fixed factor. We first present a heuristic argument, followed by formal propositions.

To derive the candidate unbalanced-growth regime, conjecture a rational expectations equilibrium for which there exist constants $k,w,p>0$ and a growth rate $G>1$ such that, as $t\to\infty$,
\begin{equation}
    K_t\sim kG^t, \quad W_t\sim wG^t, \quad P_t\sim pG^t. \label{eq:order}
\end{equation}
Suppose for the moment that the production function takes the CES form \eqref{eq:CES}. Then the land rent is given by
\begin{align*}
    r_t\coloneqq F_X(K_t,1)&=A(1-\alpha)\left(\alpha K_t^{1-\rho}+1-\alpha\right)^\frac{\rho}{1-\rho} \\
    &\sim \begin{cases*}
        A(1-\alpha)\alpha^\frac{\rho}{1-\rho}k^\rho G^{\rho t} & if $\rho<1$,\\
        A(1-\alpha)k^\alpha G^{\alpha t} & if $\rho=1$,\\
        A(1-\alpha)^\frac{1}{1-\rho} & if $\rho>1$.
    \end{cases*}
\end{align*} 
Regardless of the value of $\rho$, we have $r_t/G^t\to 0$ as $t\to\infty$.\footnote{Here $r_t$ is pure land rent. Observed real-estate rents generally combine the return to structures with the return to land. For instance, if $K=K_1+K_2$, where $K_2$ denotes real-estate structures, the corresponding measured rent is $\tilde r=K_2Y_K+XY_X$, which exceeds the pure land rent $r=XY_X$ and may grow faster. Thus substantial measured real-estate rent is consistent with a declining pure-land-rent share. See \citet{Toda2025EL}; \citet[Figure~2]{HiranoToda2025PNAS} provide evidence that the output share of land-intensive industries declines with economic development.} Hence \eqref{eq:Rt} implies $R_t\sim G$. Substituting \eqref{eq:order} into \eqref{eq:dynamics}, assuming $\bar{z}_t\to\bar{z}$, and using \eqref{eq:MPK}, we obtain
\begin{equation*}
    kG=\frac{\beta\lambda mk \int_{\bar{z}}^\infty z\diff \Phi(z)}{1-\beta(\lambda\Phi(\bar{z})+1-\lambda)}, \qquad
    \bar{z}=G/m.
\end{equation*}
Canceling $k$ and eliminating $\bar{z}$, we obtain the long-run growth rate condition
\begin{equation}
    \Psi(G/m,\lambda)=0 \label{eq:steady}
\end{equation}
for $\Psi$ in \eqref{eq:Psi}. In order for the economy to grow as conjectured, we need $G>1$. Setting $G=1$ in \eqref{eq:steady} and solving for $\lambda$ identifies the leverage threshold separating the candidate steady-state and unbalanced-growth regimes, which is the same threshold as in \eqref{eq:lambda_threshold}. The following proposition characterizes the steady-state condition and the candidate asymptotic growth rate.

\begin{prop}[Steady state and candidate growth rate]\label{prop:asymptotic}
Suppose Assumptions \ref{asmp:Phi}--\ref{asmp:F} hold and define the leverage threshold $\bar{\lambda}$ by \eqref{eq:lambda_threshold}. Then:
\begin{enumerate}
    \item\label{item:asymptotic1} If $\lambda<\bar{\lambda}$ and Assumption \ref{asmp:smallK} also holds, the dynamics \eqref{eq:dynamics} has a unique admissible steady state $(K^*,v^*)$.
    \item\label{item:asymptotic2} If $\lambda>\bar{\lambda}$, the dynamics \eqref{eq:dynamics} has no admissible steady state, and hence there is no stationary rational expectations equilibrium, but there exists a unique $G>1$ solving \eqref{eq:steady}.
\end{enumerate}
\end{prop}

Under Assumption \ref{asmp:smallK}, the aggregate dynamics \eqref{eq:dynamics} admits a unique admissible steady state at low leverage. Thus, if the initial aggregate capital $K_0$ equals the steady-state value, a rational expectations equilibrium with constant aggregate variables (stationary equilibrium) exists. The next proposition extends this existence result to nearby initial capital.

\begin{prop}[Local asymptotically balanced-growth equilibrium]\label{prop:stable}
Suppose Assumptions \ref{asmp:Phi}--\ref{asmp:smallK} hold, $\lambda<\bar\lambda$, and $(K^*,v^*)$ is the unique steady state in Proposition \ref{prop:asymptotic}\ref{item:asymptotic1}. Suppose in addition that $\Phi$ admits a continuous density $\phi$ on a neighborhood of $v^*$ with $\phi(v^*)>0$. Then, for every $K_0$ sufficiently close to $K^*$, there exists a locally unique rational expectations equilibrium satisfying $(K_t,\bar z_t)\to(K^*,v^*)$. This equilibrium is an asymptotically balanced-growth equilibrium.
\end{prop}

The density condition is a local nondegeneracy condition that makes the forward threshold equation locally invertible. It plays the same role in Proposition \ref{prop:exist} below.

We next study unbalanced-growth equilibria. The elasticity of substitution between capital and land plays a crucial role in determining the asymptotic behavior of the model.\footnote{\citet{HiranoToda2025JPE,HiranoToda2025PNAS} point out the importance of the elasticity of substitution for generating asset bubbles. See \citet{Sorger2026} for an analysis of an overlapping generations model with endogenous growth.} The relevant object is the elasticity associated with $F$, not necessarily that associated with the production function $Y$ (which is $1/\rho$ in the CES case \eqref{eq:CES}). For $F$, define the elasticity as the percentage change in relative factor inputs with respect to the percentage change in relative factor prices:
\begin{equation}
    \sigma(K,X)=-\frac{\partial \log (K/X)}{\partial \log (F_K/F_X)}. \label{eq:sigma}
\end{equation}

\begin{asmp}\label{asmp:ES}
The elasticity of substitution between capital and land defined by \eqref{eq:sigma} exceeds 1 at high capital levels: $\sigma\coloneqq \liminf_{K\to\infty}\sigma(K,1)>1$.
\end{asmp}

Empirical estimates of the elasticity of substitution between land and structures in housing production are typically above 1 \citep{EppleGordonSieg2010,AhlfeldtMcMillen2018}. Although these estimates concern production functions such as $Y$ rather than the composite function $F$, they are consistent with the CES case $\rho<1$. The calculation below shows that, because $F$ includes undepreciated capital, the CES specification \eqref{eq:CES} satisfies Assumption \ref{asmp:ES} for every $\rho>0$. Suppose first that $\rho\neq1$ and define
\begin{equation*}
    S(K)\coloneqq \alpha K^{1-\rho}+1-\alpha.
\end{equation*}
At $X=1$, direct differentiation gives
\begin{align*}
    F_K(K,1)&=1-\delta+A\alpha K^{-\rho}S(K)^{\rho/(1-\rho)},\\
    F_X(K,1)&=A(1-\alpha)S(K)^{\rho/(1-\rho)},
\end{align*}
and hence
\begin{equation*}
    \frac{1}{\sigma(K,1)}
    =\frac{\diff\log (F_X/F_K)(K,1)}{\diff\log K}
    =\frac{\rho\alpha K^{1-\rho}}{S(K)}
    +\frac{\rho(1-\alpha)}{S(K)}\frac{Y_K(K,1)}{F_K(K,1)}.
\end{equation*}
It follows that $1/\sigma(K,1)\to\rho$ if $\rho<1$ and $1/\sigma(K,1)\to0$ if $\rho>1$. If $\rho=1$, then $Y_K(K,1)=A\alpha K^{\alpha-1}$ and $F_X(K,1)=A(1-\alpha)K^\alpha$, so
\begin{equation*}
    \frac{1}{\sigma(K,1)}
    =\alpha+(1-\alpha)\frac{Y_K(K,1)}{F_K(K,1)}\to\alpha.
\end{equation*}
Therefore,
\begin{equation*}
    \lim_{K\to\infty} \sigma(K,1)=\begin{cases*}
        1/\rho & if $\rho<1$,\\
        1/\alpha & if $\rho=1$,\\
        \infty & if $\rho>1$,
    \end{cases*}
\end{equation*}
which exceeds 1 for all parameter values. More generally, the strict inequality in Assumption \ref{asmp:ES} implies that the land share and the deviation of the average product of capital from $m$ decay at a polynomial rate as capital grows. These rate bounds allow us to establish convergence after detrending by $G$, rather than only convergence of one-period growth rates. The following proposition establishes that the candidate rate is attained by a locally unique unbalanced-growth equilibrium.

\begin{prop}[Local unbalanced-growth equilibrium]\label{prop:exist}
Suppose Assumptions \ref{asmp:Phi}--\ref{asmp:F} and \ref{asmp:ES} hold and let $\bar{\lambda}$ be the leverage threshold in \eqref{eq:lambda_threshold}. Suppose $\lambda>\bar{\lambda}$, let $G>1$ be the unique growth rate in Proposition \ref{prop:asymptotic}\ref{item:asymptotic2}, and set $\bar{z}=G/m$. Suppose in addition that $\bar{z}$ is admissible and that $\Phi$ admits a continuous density $\phi$ on a neighborhood of $\bar{z}$ with $\phi(\bar{z})>0$. Then, for any sufficiently large $K_0>0$, there exists a locally unique rational expectations equilibrium satisfying $(K_t,\bar{z}_t)\to(\infty,\bar{z})$ and $(K_t,W_t,P_t)/G^t\to (k,w,p)$ for some constants $k,w,p>0$.
\end{prop}

Here local uniqueness is understood in the sequence-space sense made precise in the proof. Using $\Psi(\bar{z},\lambda)=0$, we can rewrite the admissibility of the limiting threshold as $\beta\E[Z\mid Z\ge\bar{z}]>\bar{z}$. This is exactly the conditional tail expectation criterion in Proposition \ref{prop:admissible_leverage}\ref{item:admissible_iff}: average productivity among agents above the investment cutoff must be sufficiently high relative to the cutoff. Thus the same upper-tail mechanism identified in \S\ref{sec:loose_constraints} governs whether the candidate growth path satisfies the equilibrium conditions in the general model. The condition also ensures that the limiting land-price coefficient $p$ is positive.

Proposition \ref{prop:exist} revisits the financial accelerator described above. Along the locally constructed unbalanced-growth equilibrium, rising land prices raise wealth and investment, while capital accumulation feeds back into wealth and weakly raises future land rents. This feedback sustains the common asymptotic growth rate $G$ of capital, wealth, and land prices.

Why can endogenous rents not support the higher land price? Under Assumption \ref{asmp:ES}, Lemma \ref{lem:MRT} implies that $r_t=F_X(K_t,1)=O(K_t^\nu)$ for some $\nu\in(0,1)$. Proposition \ref{prop:exist} implies that both $K_t$ and $P_t$ grow asymptotically at rate $G>1$, so $r_t/P_t=O(G^{-(1-\nu)t})$ and hence $\sum_{t=1}^\infty r_t/P_t<\infty$. By the Bubble Characterization Lemma in \eqref{eq:bubble_characterization}, this summability implies $P_t>V_t$ at every date. Thus the endogenous rent response is too weak to account for the land price: the remainder is a bubble. Combining the preceding results yields the following asset-pricing characterization within the two long-run regimes considered here.

\begin{thm}[Growth regimes and land bubbles: general model]\label{thm:land_bubble}
Suppose Assumptions \ref{asmp:Phi}--\ref{asmp:F} hold and let $\bar{\lambda}$ be as in \eqref{eq:lambda_threshold}. Then:
\begin{enumerate}
    \item\label{item:land_bubble_balanced} If $\lambda<\bar{\lambda}$,  Assumption \ref{asmp:smallK} holds, and the additional density condition in Proposition \ref{prop:stable} holds, its equilibrium exhibits asymptotically balanced growth and has no land bubble.
    \item\label{item:land_bubble_unbalanced} If $\lambda>\bar{\lambda}$, Assumption \ref{asmp:ES} holds, and the additional admissibility and density conditions in Proposition \ref{prop:exist} hold, its equilibrium exhibits unbalanced growth and a land bubble.
\end{enumerate}
\end{thm}

Theorem \ref{thm:land_bubble} is an asset-pricing characterization within the two long-run regimes stated in its two parts. It neither rules out other nonstationary equilibria nor establishes convergence from arbitrary initial capital.

\subsection{Numerical illustration}\label{sec:discuss}

We use the general model to illustrate how a change in leverage affects capital, land prices, rents, and the price--rent ratio. The exercise is deliberately confined to leverage, the parameter at the center of the paper's comparative-static result, and is neither a calibration exercise nor an empirical test of whether an observed change in credit conditions caused a bubble. Under the exponential productivity distribution \eqref{eq:exp}, the leverage threshold \eqref{eq:lambda_threshold} equals
\begin{equation}
    \bar{\lambda}=\frac{1-\beta}{\beta}\frac{\gamma}{m}\e^\frac{\gamma}{m}. \label{eq:lambda_exmp}
\end{equation}

For this specification, a derivation in Appendix \ref{sec:num} yields
\begin{equation}
    \int_{\bar{z}}^\infty z\diff \Phi(z)=(\bar{z}+1/\gamma)\e^{-\gamma\bar{z}}. \label{eq:zint}
\end{equation}
Therefore \eqref{eq:dynamics_K} simplifies to
\begin{equation}
    K_{t+1}=\frac{\beta\lambda(\bar{z}_t+1/\gamma)\e^{-\gamma\bar{z}_t}}{1-\beta+\beta\lambda \e^{-\gamma\bar{z}_t}}F(K_t,1). \label{eq:dynamics_K1}
\end{equation}
We numerically solve the model using the algorithm discussed in Appendix \ref{sec:num}. We set $A=1$, $\alpha=0.5$, and $\rho=1$ in the CES production function \eqref{eq:CES} and $\delta=0.08$, $\beta=0.95$, and $\gamma=-\log(0.1)$. Thus $m=0.92$, and \eqref{eq:lambda_exmp} gives $\bar\lambda\approx1.61$.

\begin{figure}[htb!]
\centering
\includegraphics[width=0.7\linewidth]{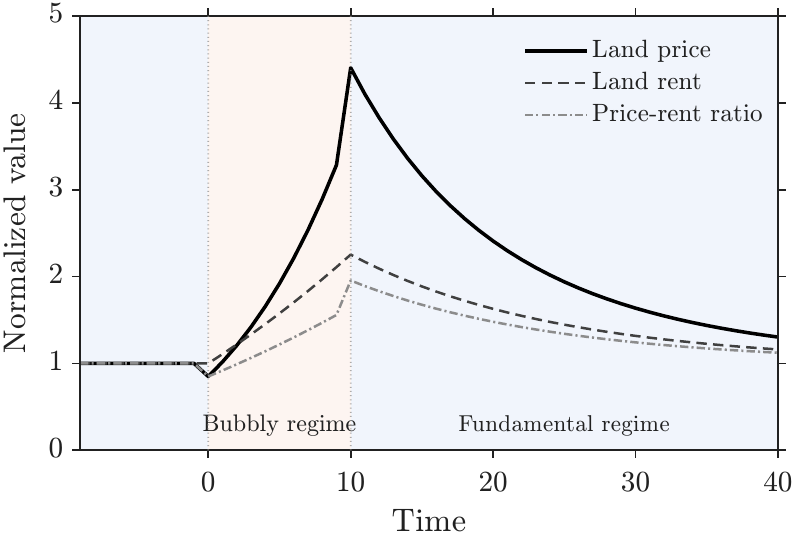}
\caption{Numerical transition under changes in leverage. Leverage unexpectedly rises from $\lambda=1$ to $\lambda=2$ at $t=0$ and unexpectedly returns to $\lambda=1$ at $t=10$. Variables are normalized to one in the initial steady state.}\label{fig:dynamics}
\end{figure}

Figure~\ref{fig:dynamics} presents a simple sequence of parameter experiments. The economy initially rests at the fundamental steady state with $\lambda=1$. At $t=0$, leverage unexpectedly rises to $\lambda=2>\bar\lambda$. Conditional on the information then available, agents expect leverage to remain at its new value. Capital and land prices subsequently grow, and the price--rent ratio rises because the land price grows faster than rent. At $t=10$, leverage unexpectedly returns to $\lambda=1<\bar\lambda$; land prices, rents, and the price--rent ratio then decline. At the instant leverage changes, output is predetermined. Hence the loosening at $t=0$ shifts saving from land toward capital and initially depresses the land price, whereas the tightening at $t=10$ reverses this substitution and initially raises it. Away from the two change dates, the paths satisfy perfect foresight conditional on the prevailing leverage level. Thus, the changes at $t=0$ and $t=10$ are MIT shocks rather than realizations of an explicitly modeled stochastic process. The exercise visualizes a numerical transition associated with crossing the threshold in both directions; it is not a claim that an observed credit reform is exogenous or a theorem about global transition dynamics.

Japan's land-price boom in the 1980s provides useful motivation for studying leverage and land valuation jointly: financial liberalization, credit expansion, and rapidly rising land prices occurred together. The episode does not, however, provide a clean exogenous change in the model's leverage parameter. Nor is the numerical exercise calibrated to distinguish a leverage-driven boom from alternative explanations or to establish that a particular regulatory or monetary intervention caused the subsequent decline in land prices. We therefore view Japan as motivation rather than evidence for the model's causal mechanism; see \citet[\S1.1]{HiranoToda2024JME} and the references therein for a fuller historical discussion.

\section{Concluding remarks}

We show that relaxing leverage can make a land bubble necessary, not merely possible. Below the threshold, each leverage level has a unique capital level that supports balanced growth, and land is priced at the present value of rents. Above the threshold, whenever an equilibrium exists, the fundamental regime disappears and the unique equilibrium in the class we consider exhibits unbalanced growth: demand for a store of value makes land prices grow faster than rents. The upper tail of the productivity distribution determines how far this bubble necessity regime extends. Bounded or sufficiently thin tails eventually rule it out, whereas a sufficiently thick unbounded tail sustains it at arbitrarily high leverage. The general model shows that the mechanism survives when the unmodeled financial sector is removed and rents are endogenous.

\appendix
\section{Proofs}

\subsection{Proof of Theorem \ref{thm:trend}}

We prove Theorem \ref{thm:trend} by establishing a series of lemmas. Below, Assumptions \ref{asmp:Phi} and \ref{asmp:bmz} are in force.

\begin{lem}[No contracting equilibrium]\label{lem:G<1}
In any equilibrium in Definition \ref{defn:lr}, we have $G\ge 1$.
\end{lem}

\begin{proof}
Iterating \eqref{eq:noarbitrage_R} forward and using nonnegative land prices shows that $P_t\ge V=D/(R-1)$. If $G<1$, then \eqref{eq:Wt1} implies $W_t\to0$. Let
\begin{equation*}
    b\coloneqq\beta(\lambda\Phi(\bar z)+1-\lambda)>0.
\end{equation*}
Equation \eqref{eq:Pt1} gives $N_t=bW_t-P_t\le bW_t-V$. Hence $\abs{N_t/W_t}\to\infty$, contradicting condition \ref{item:N/W}. Therefore $G\ge1$.
\end{proof}

\begin{lem}[Structure of balanced growth]\label{lem:G=1}
In an equilibrium with $G=1$, we have $N_t=0$ and $P_t=V=D/(R-1)$.
\end{lem}

\begin{proof}
If $G=1$, aggregate wealth is constant, so condition \ref{item:N/W} implies $N_t\to0$. Equation \eqref{eq:Pt1} then implies that $P_t$ converges. The general solution to \eqref{eq:noarbitrage_R} is $P_t=V+C R^t$. Because $R>1$ and $P_t$ converges, $C=0$ and hence $P_t=V$. Equation \eqref{eq:Pt1} now makes $N_t$ constant, so $N_t=0$.
\end{proof}

\begin{lem}[Structure of unbalanced growth]\label{lem:G>1}
In an unbalanced-growth equilibrium, $G=R$, $N_t=-D/(R-1)$, and $P_t$ satisfies \eqref{eq:Pt_b}.
\end{lem}

\begin{proof}
Let $b\coloneqq\beta(\lambda\Phi(\bar z)+1-\lambda)>0$. Since $W_t=G^tW_0$ and $N_t/W_t\to0$, \eqref{eq:Pt1} implies $P_t/W_t\to b$. Therefore
\begin{equation*}
    R=\frac{P_{t+1}+D}{P_t}
    =G\frac{P_{t+1}/W_{t+1}+D/W_{t+1}}{P_t/W_t}\to G,
\end{equation*}
so $G=R$. Iterating \eqref{eq:noarbitrage_R} forward gives
\begin{equation}
    P_t=\sum_{s=1}^TR^{-s}D+R^{-T}P_{t+T}. \label{eq:Pt_iter}
\end{equation}
On the other hand, \eqref{eq:Pt1} and $G=R$ imply $P_t+N_t=bW_t=bR^tW_0$, so
\begin{equation}
    P_t+N_t=R^{-T}(P_{t+T}+N_{t+T}). \label{eq:PNt_iter}
\end{equation}
Subtracting \eqref{eq:Pt_iter} from \eqref{eq:PNt_iter}, we obtain
\begin{equation}
    N_t=-D\sum_{s=1}^TR^{-s}+R^{-T}N_{t+T}. \label{eq:Nt_iter}
\end{equation}
Because $W_t=R^tW_0$, condition \ref{item:N/W} gives $R^{-T}N_{t+T}\to0$. Letting $T\to\infty$ in \eqref{eq:Nt_iter} yields $N_t=-V$. Hence $P_t=bW_t+V=bW_0R^t+V$. Using $P_0=bW_0+V$ and $V+D=RV$, the initial balance sheet in Definition \ref{defn:lr} gives
\begin{equation*}
    W_0=mK_0+bW_0+RV,
    \qquad
    W_0=\frac{mK_0+RV}{1-b}.
\end{equation*}
Substitution yields \eqref{eq:Pt_b}.
\end{proof}

\begin{proof}[Proof of Theorem \ref{thm:trend}]
By Lemma \ref{lem:G<1}, only $G=1$ and $G>1$ need be considered.

Suppose first that $G=1$. By \eqref{eq:G}, write the economic growth rate as $G(R;\lambda)\coloneqq\beta(R+\lambda\pi(R))$. For $\lambda<\bar\lambda$ and $R\ge1$, Assumption \ref{asmp:bmz} and admissibility at the threshold imply
\begin{equation*}
    \lambda[1-\Phi(R/m)]
    \le\lambda[1-\Phi(1/m)]<1.
\end{equation*}
Consequently,
\begin{equation*}
    G_R(R;\lambda)=\beta[\lambda\Phi(R/m)+1-\lambda]>0.
\end{equation*}
By \eqref{eq:return_growth_gap} and \eqref{eq:Psi_threshold}, $G(1;\lambda)<1$. Moreover, Assumption \ref{asmp:bmz} implies $\pi(1/\beta)>0$, so $G(1/\beta;\lambda)>1$. Thus there is a unique $R\in(1,1/\beta)$ satisfying $G(R;\lambda)=1$. Lemma \ref{lem:G=1} gives $P_t=V$ and $N_t=0$. Writing $b=\beta(\lambda\Phi(\bar z)+1-\lambda)$, \eqref{eq:Pt1} uniquely determines $W_t=V/b$. The initial balance sheet becomes $W_0=mK_0+RV$. Because $1=Rb+\beta\lambda m\int_{R/m}^\infty z\diff\Phi(z)$, this condition holds if and only if \eqref{eq:K0_balanced} holds. In that case, \eqref{eq:Kt+} gives $K_{t+1}=K_0$ for all $t\ge0$. This constructs the unique balanced-growth equilibrium. Its no-bubble condition holds because $R^{-T}P_{t+T}=R^{-T}V\to0$.

Conversely, any balanced-growth equilibrium must have $R<1/\beta$ and satisfy
\begin{equation*}
    \frac{1}{\lambda}=\frac{\beta\pi(R)}{1-\beta R}.
\end{equation*}
The expression on the right is strictly increasing on $[1,1/\beta)$. Indeed, the sign of its derivative is the sign of
\begin{equation*}
    \Phi(R/m)-1+\beta m\int_{R/m}^\infty z\diff\Phi(z),
\end{equation*}
which is positive at $R=1$ by Assumption \ref{asmp:bmz} and weakly increasing on this interval. Its value at $R=1$ is $1/\bar\lambda$. Hence a balanced-growth equilibrium requires $1/\lambda>1/\bar\lambda$, or $\lambda<\bar\lambda$. Lemma \ref{lem:G=1}, \eqref{eq:Pt1}, and the initial balance sheet then imply \eqref{eq:K0_balanced}.

If $\lambda\le\bar\lambda$, no unbalanced-growth equilibrium exists: $R>1$ implies $R/m>1/m$, while \eqref{eq:Psi_threshold} and strict monotonicity give $\Psi(R/m,\lambda)>0$, contradicting $G=R$. If $\lambda>\bar\lambda$, an unbalanced-growth equilibrium must satisfy $G=R$ by Lemma \ref{lem:G>1}. By \eqref{eq:return_growth_gap}, this is equivalent to
\begin{equation*}
    \Psi(R/m,\lambda)=0.
\end{equation*}
Because $\Psi(1/m,\lambda)<0$, $\Psi_v>0$, and $\Psi(v,\lambda)>0$ for all sufficiently large $v$, there is a unique solution $R=mv_\lambda>1$. It defines an equilibrium if and only if $v_\lambda$ is admissible. To verify sufficiency, let
\begin{equation*}
    b\coloneqq\beta(\lambda\Phi(v_\lambda)+1-\lambda),
    \quad W_0\coloneqq\frac{mK_0+RV}{1-b},
    \quad W_t=R^tW_0,
\end{equation*}
set
\begin{equation*}
    K_{t+1}\coloneqq\beta\lambda W_t\int_{v_\lambda}^\infty z\diff\Phi(z),
    \qquad
    N_t\coloneqq-V,
    \qquad
    P_t\coloneqq bW_t+V.
\end{equation*}
Because $P_0+D=bW_0+RV$, the definition of $W_0$ verifies the initial balance sheet. The equation $G=R$ implies
\begin{equation*}
    R(1-b)=\beta\lambda m\int_{v_\lambda}^\infty z\diff\Phi(z),
\end{equation*}
and all remaining equilibrium conditions follow directly, including $N_t/W_t\to0$. Lemma \ref{lem:G>1} establishes uniqueness. Moreover, \eqref{eq:Pt_b} implies
\begin{equation*}
    \lim_{T\to\infty}R^{-T}P_{t+T}=bW_0R^t>0,
\end{equation*}
so the no-bubble condition fails and the equilibrium is bubbly.

Finally, if $\lambda=\bar\lambda$, the only candidate separating the two equations has $G=R=1$. It belongs to neither equilibrium class and violates the maintained requirement $R>1$. This proves the third statement.
\end{proof}

\subsection{Proof of Proposition \ref{prop:rate}}

In the balanced-growth region, \eqref{eq:G} gives the economic growth rate as $G(R;\lambda)\coloneqq\beta(R+\lambda\pi(R))$. The equilibrium condition is $G(R;\lambda)=1$. Because $\pi'(R)=-(1-\Phi(R/m))$, we have
\begin{equation*}
    G_R(R;\lambda)=\beta[\lambda\Phi(R/m)+1-\lambda]>0,
    \qquad
    G_\lambda(R;\lambda)=\beta\pi(R)>0.
\end{equation*}
By the implicit function theorem, we have $\diff R/\diff\lambda<0$. In the unbalanced-growth region, $\Psi(v_\lambda,\lambda)=0$ and
\begin{equation*}
    \frac{\diff v_\lambda}{\diff\lambda}
    =\frac{\beta\int_{v_\lambda}^\infty(z-v_\lambda)\diff\Phi(z)}{\Psi_v(v_\lambda,\lambda)}>0.
\end{equation*}
Since $R=mv_\lambda$, the second claim follows. Continuity and $\Psi(1/m,\bar\lambda)=0$ imply the limit at the threshold. \hfill \qedsymbol

\subsection{Proof of Proposition \ref{prop:asymptotic}}

We prove Proposition \ref{prop:asymptotic} by establishing two lemmas. Below, Assumptions \ref{asmp:Phi}--\ref{asmp:F} are in force. If an admissible steady state $(K,v)$ exists, then \eqref{eq:WPKz}, \eqref{eq:dynamics} imply
\begin{subequations}
	\begin{align}
		K&=\frac{\beta\lambda F}{1-\beta(\lambda\Phi(v)+1-\lambda)}\int_v^\infty z\diff \Phi(z), \label{eq:steady_K}\\
		v&=\frac{1}{F_K}\left(1+\frac{1-\beta(\lambda\Phi(v)+1-\lambda)}{\beta(\lambda\Phi(v)+1-\lambda)}\frac{F-KF_K}{F}\right), \label{eq:steady_z}
	\end{align}
\end{subequations}
where $F,F_K$ are evaluated at $(K,1)$ and we have used $F(K,1)=KF_K(K,1)+F_X(K,1)$.

\begin{lem}[Stationary capital conditional on cutoff]\label{lem:steady_K}
Suppose Assumption \ref{asmp:smallK} holds and $\lambda<\bar{\lambda}$. Let $v$ be admissible with $\Phi(v)<1$. Then there exists a unique $K>0$ solving \eqref{eq:steady_K} if and only if
\begin{equation}
    \frac{1-\beta(\lambda\Phi(v)+1-\lambda)}{\beta\lambda \int_v^\infty z\diff\Phi(z)}<\bar{m}. \label{eq:steady_K_upper}
\end{equation}
Moreover, the set of thresholds satisfying \eqref{eq:steady_K_upper} has a connected component containing $1/m$.
\end{lem}

\begin{proof}
We can rewrite \eqref{eq:steady_K} as
\begin{equation}
    \frac{1-\beta(\lambda\Phi(v)+1-\lambda)}{\beta\lambda \int_v^\infty z\diff\Phi(z)}=\frac{F(K,1)}{K}. \label{eq:steady_K2}
\end{equation}
By \eqref{eq:MPK} and \eqref{eq:mbar}, the right-hand side of \eqref{eq:steady_K2} is strictly decreasing from $\bar{m}$ to $m$ as $K$ increases from 0 to $\infty$. Hence \eqref{eq:steady_K2} has a unique solution if and only if its left-hand side lies in $(m,\bar{m})$. We first show that the lower inequality always holds. To this end, observe
\begin{subequations}
\begin{align}
    &\frac{1-\beta(\lambda\Phi(v)+1-\lambda)}{\beta\lambda \int_v^\infty z\diff\Phi(z)}>m \label{eq:steady_K2_lower}\\
    \iff &1-\beta(\lambda\Phi(v)+1-\lambda)-\beta\lambda m\int_v^\infty z\diff \Phi(z)>0. \label{eq:steady_K2_equiv}
\end{align}
\end{subequations}
Writing $\phi=\Phi'\ge 0$, since the derivative of the left-hand side of \eqref{eq:steady_K2_equiv} with respect to $v$ is $\beta\lambda\phi(v)(mv-1)$ almost everywhere, it is decreasing for $v<1/m$ and increasing for $v>1/m$. It therefore achieves a minimum at $v=1/m$, where its value is $m\Psi(1/m,\lambda)>0$ by \eqref{eq:Psi_threshold}. Hence \eqref{eq:steady_K2_lower} holds for every admissible $v$. It thus remains to establish the upper inequality \eqref{eq:steady_K_upper}. Assumption \ref{asmp:bmz} implies $\Phi(1/m)<1$ and
\begin{equation*}
    \beta m\int_{1/m}^\infty z\diff\Phi(z)>1-\Phi(1/m).
\end{equation*}
Therefore, using the definition \eqref{eq:lambda_threshold},
\begin{equation*}
    \bar{\lambda}\left(1-\Phi(1/m)\right)
    =\frac{(1-\beta)(1-\Phi(1/m))}
    {\beta\left(m\int_{1/m}^\infty z\diff\Phi(z)-(1-\Phi(1/m))\right)}<1.
\end{equation*}
Because $\lambda<\bar{\lambda}$, it follows that $\lambda(1-\Phi(1/m))<1$, or equivalently $\lambda\Phi(1/m)+1-\lambda>0$. Thus $1/m$ is admissible. At this threshold, the left-hand side of \eqref{eq:steady_K2} is
\begin{align*}
    \frac{1-\beta(\lambda\Phi(1/m)+1-\lambda)}{\beta\lambda \int_{1/m}^\infty z\diff\Phi(z)}
    &=\frac{1-\beta}{\beta\lambda\int_{1/m}^\infty z\diff\Phi(z)}
      +\frac{1-\Phi(1/m)}{\int_{1/m}^\infty z\diff\Phi(z)}\\
    &\le \frac{1-\beta\Phi(1/m)}{\beta\int_{1/m}^\infty z\diff\Phi(z)}<\bar{m},
\end{align*}
where the first inequality uses $\lambda\ge 1$ and the second inequality is equivalent to Assumption \ref{asmp:smallK}. Therefore $1/m$ satisfies \eqref{eq:steady_K_upper}; continuity yields a connected component containing it.
\end{proof}

When $\bar{m}<\infty$, condition \eqref{eq:steady_K_upper} necessarily fails for thresholds sufficiently close to the upper endpoint of the productivity distribution because its left-hand side tends to infinity. Thus, even when $\lambda<\bar{\lambda}$, Lemma \ref{lem:steady_K} does not assert that \eqref{eq:steady_K} has a solution for every admissible threshold. The lemma also deliberately makes no claim when $\lambda\ge\bar{\lambda}$: in that case \eqref{eq:steady_K} may have a solution for some values of $v$. Ruling out an admissible steady state requires imposing \eqref{eq:steady_z} as well.

\begin{lem}[Existence and uniqueness of steady state]\label{lem:steady_exist}
If $\lambda\ge\bar{\lambda}$, the dynamics \eqref{eq:dynamics} has no admissible steady state. If Assumption \ref{asmp:smallK} also holds and $\lambda<\bar{\lambda}$, it has a unique admissible steady state.
\end{lem}

\begin{proof}
Suppose first that $\lambda\ge\bar{\lambda}$ and, to obtain a contradiction, that $(K,v)$ is an admissible steady state. For this calculation only, write
\begin{equation*}
    b\coloneqq\beta(\lambda\Phi(v)+1-\lambda)\in (0,\beta),
    \qquad I\coloneqq\int_v^\infty z\diff\Phi(z)>0.
\end{equation*}
Equation \eqref{eq:steady_K} implies
\begin{equation*}
    \frac{K}{F}=\frac{\beta\lambda I}{1-b},
    \qquad
    \frac{F-KF_K}{F}=1-\frac{\beta\lambda F_KI}{1-b}.
\end{equation*}
Substituting the second equality into \eqref{eq:steady_z}, we obtain
\begin{equation*}
    v=\frac{1}{F_K}\left[1+\frac{1-b}{b}
        \left(1-\frac{\beta\lambda F_KI}{1-b}\right)\right]=\frac{1}{F_K}\left(\frac{1}{b}
        -\frac{\beta\lambda F_KI}{b}\right).
\end{equation*}
Multiplying by $b$ and rearranging yields
\begin{align}
    \frac{1}{F_K(K,1)}
    &=bv+\beta\lambda I\notag \\
	&=\beta\left((\lambda\Phi(v)+1-\lambda)v+\lambda\int_v^\infty z\diff\Phi(z)\right)
    =v-\Psi(v,\lambda), \label{eq:steady_Psi}
\end{align}
where the last equality follows from \eqref{eq:Psi} and
\begin{equation*}
    \int_v^\infty(z-v)\diff\Phi(z)
    =I-v(1-\Phi(v)).
\end{equation*}
At a steady state, \eqref{eq:Rt} and \eqref{eq:z_threshold} imply
\begin{equation*}
    F_K(K,1)v=\frac{P+F_X(K,1)}{P}>1,
\end{equation*}
and hence \eqref{eq:steady_Psi} and \eqref{eq:Psi_threshold} imply
\begin{equation*}
    \Psi(v,\lambda)>0\ge\Psi(1/m,\lambda).
\end{equation*}
Because $\Psi(\cdot,\lambda)$ is strictly increasing by \eqref{eq:Psi_v}, it follows that $v>1/m$.

By \eqref{eq:Psi_v}, the derivative of $v\mapsto v-\Psi(v,\lambda)$ is $\beta(\lambda\Phi(v)+1-\lambda)$. If $\lambda\Phi(1/m)+1-\lambda>0$, this derivative is positive on $[1/m,v]$, so
\begin{equation*}
    v-\Psi(v,\lambda)>1/m-\Psi(1/m,\lambda)\ge 1/m.
\end{equation*}
If $\lambda\Phi(1/m)+1-\lambda\le 0$, choose $\ubar{v}\ge 1/m$ such that $\Phi(\ubar{v})=1-1/\lambda$. Because the steady state is admissible, $v>\ubar{v}$, and $v-\Psi(v,\lambda)$ is increasing on $[\ubar{v},v]$. Furthermore, using Assumption \ref{asmp:bmz}, \eqref{eq:Psi}, and $\Phi(\ubar{v})=1-1/\lambda$, we have
\begin{align*}
    \ubar{v}-\Psi(\ubar{v},\lambda)
    &=\beta\lambda\int_{\ubar{v}}^\infty z\diff\Phi(z)
      =\beta\E[Z\mid Z\ge \ubar{v}]\\
    &\ge \beta\E[Z\mid Z\ge 1/m]>1/m.
\end{align*}
Thus, in either case, \eqref{eq:steady_Psi} implies $1/F_K(K,1)>1/m$. But concavity of $F$ and \eqref{eq:MPK} imply $F_K(K,1)\ge m$, a contradiction. Therefore no admissible steady state exists when $\lambda\ge\bar{\lambda}$.

Let Assumption \ref{asmp:smallK} hold and $\lambda<\bar{\lambda}$. For this proof only, let $\ell(v)$ denote the left-hand side of \eqref{eq:steady_K2} as a function of the candidate threshold $v$. By Lemma \ref{lem:steady_K}, the set of admissible $v$ satisfying $\Phi(v)<1$ and $\ell(v)<\bar{m}$ contains $1/m$, and for each $v$ in this set there exists a unique $K(v)>0$ solving \eqref{eq:steady_K}; moreover, $K(v)$ is continuous on every connected component. Because $F(K,1)/K$ is strictly decreasing in $K$, $K(v)$ is decreasing on any interval on which $\ell(v)$ is increasing. Let $\cI=(v_-,v_+)$ be the component containing $1/m$. Given \eqref{eq:steady_K}, using \eqref{eq:steady_Psi}, equation \eqref{eq:steady_z} is equivalent to
\begin{equation}
    \frac{1}{F_K(K(v),1)}=v-\Psi(v,\lambda). \label{eq:steady_Psi_v}
\end{equation}

Writing $\phi=\Phi'$ and differentiating $\ell(v)$, we obtain
\begin{equation}
    \ell'(v)=\frac{\phi(v)}{\int_v^\infty z\diff\Phi(z)}(v\ell(v)-1) \label{eq:ell_prime}
\end{equation}
almost everywhere on the domain of $\ell$. Using the expression for $v-\Psi(v,\lambda)$ in \eqref{eq:steady_Psi} and the definition of $\ell(v)$, we obtain
\begin{equation}
    \ell(v)(v-\Psi(v,\lambda))-1
    =\beta(\lambda\Phi(v)+1-\lambda)(v\ell(v)-1). \label{eq:ell_Psi}
\end{equation}
Euler's theorem and $F_X>0$ also imply
\begin{equation}
    F_K(K(v),1)<\frac{F(K(v),1)}{K(v)}=\ell(v). \label{eq:FK_ell}
\end{equation}
For every $v$ for which $K(v)$ exists, define the steady-state residual
\begin{equation*}
    H(v)\coloneqq \frac{1}{F_K(K(v),1)}-(v-\Psi(v,\lambda)).
\end{equation*}
Thus \eqref{eq:steady_Psi_v} holds if and only if $H(v)=0$. If $v\ell(v)\le 1$, then \eqref{eq:ell_Psi} implies $v-\Psi(v,\lambda)\le 1/\ell(v)$, whereas \eqref{eq:FK_ell} implies $1/F_K(K(v),1)>1/\ell(v)$. Hence $H(v)>0$, so any steady-state threshold must satisfy $v\ell(v)>1$.

We next characterize the set of such candidate thresholds. The sign of $v\ell(v)-1$ is the same as that of
\begin{equation*}
    v\set*{1-\beta(\lambda\Phi(v)+1-\lambda)}
    -\beta\lambda\int_v^\infty z\diff\Phi(z),
\end{equation*}
whose derivative almost everywhere is $1-\beta(\lambda\Phi(v)+1-\lambda)>0$. Therefore the region where $v\ell(v)>1$ is an interval. Furthermore, Lemma \ref{lem:steady_K} implies $\ell(1/m)>m$, and hence $\ell(1/m)/m>1$. On the region where $v\ell(v)>1$, \eqref{eq:ell_prime} implies that $\ell$ is increasing. Consequently, among admissible thresholds satisfying $\Phi(v)<1$, the conditions $v\ell(v)>1$ and $\ell(v)<\bar{m}$ define a single interval
\begin{equation*}
    \cJ\coloneqq(v_0,v_+)\subset\cI
\end{equation*}
containing $1/m$. By Lemma \ref{lem:steady_K} and the preceding exclusion, every possible steady-state threshold lies in $\cJ$.

We now show uniqueness. Because $\ell$ is increasing on $\cJ$, the monotonicity property of $K(v)$ noted above implies that $K(v)$ is decreasing on $\cJ$. Concavity of $F$ then implies that $F_K(K(v),1)$ is increasing, so $1/F_K(K(v),1)$ is decreasing. On the other hand, differentiating the expression in \eqref{eq:steady_Psi} and using \eqref{eq:Psi_v}, we obtain
\begin{equation*}
    \frac{\diff}{\diff v}(v-\Psi(v,\lambda))
    =\beta(\lambda\Phi(v)+1-\lambda)>0.
\end{equation*}
Thus $H$ is strictly decreasing on $\cJ$, so there can be at most one steady state.

It remains to establish existence by checking the signs of $H$ at the endpoints of $\cJ$. Suppose first that $v_0>v_-$. Then $v_0\in\cI$ and $v_0\ell(v_0)=1$, so \eqref{eq:ell_Psi} and \eqref{eq:FK_ell} imply $H(v_0)>0$. Suppose instead that $v_0=v_-$. Then $v\ell(v)>1$ throughout $\cI$. Because $\ell$ is increasing and $\ell(1/m)<\bar{m}$, the lower endpoint cannot be generated by $\ell(v)$ reaching $\bar{m}$. Nor can $v_-=0$, because $\ell(v)$ has a finite limit as $v\downarrow0$ and hence $v\ell(v)-1\to-1$. It follows that $v_-$ is a boundary of the admissible region, where $\lambda\Phi(v)+1-\lambda\downarrow0$. Let $\ell_-\coloneqq\lim_{v\downarrow v_-}\ell(v)$. Because $\ell$ is increasing and $1/m\in\cI$, we have $\ell_-\le\ell(1/m)<\bar{m}$; the uniform lower bound in the proof of Lemma \ref{lem:steady_K} also gives $\ell_->m$. Hence $K(v)\to K_-\in(0,\infty)$ as $v\downarrow v_-$. Taking limits in \eqref{eq:ell_Psi} and using \eqref{eq:FK_ell}, we obtain
\begin{equation*}
    \lim_{v\downarrow v_-}H(v)
    =\frac{1}{F_K(K_-,1)}-\frac{1}{\ell_-}>0.
\end{equation*}
Thus in either case $H$ is positive near $v_0$.

It remains to obtain the opposite inequality near $v_+$. Suppose first that $\bar{m}<\infty$. Because $\ell$ is increasing to the right of $1/m$ and tends to infinity at the upper endpoint of the productivity distribution, we have $\ell(v)\to\bar{m}$ and $K(v)\to0$ as $v\uparrow v_+$. To identify the limiting marginal product, note that \eqref{eq:mbar} implies $F(K,1)\to0$ as $K\downarrow0$. Extending $K\mapsto F(K,1)$ continuously to zero, the secant-slope property of concave functions gives
\begin{equation*}
    \lim_{K\downarrow0}F_K(K,1)
    =\lim_{K\downarrow0}\frac{F(K,1)}{K}=\bar{m}.
\end{equation*}
Therefore $F_K(K(v),1)\to\bar{m}$. Since $v_+>1/m$ and $\bar{m}>m$, we have $v_+\bar{m}>1$, so \eqref{eq:ell_Psi} implies
\begin{equation*}
    v_+-\Psi(v_+,\lambda)>1/\bar{m}
    =\lim_{v\uparrow v_+}\frac{1}{F_K(K(v),1)}.
\end{equation*}
Hence $H(v)<0$ near $v_+$.

Finally, suppose $\bar{m}=\infty$, and let $z_{\max}\in(0,\infty]$ be the upper endpoint of the support of $\Phi$. Then $v_+=z_{\max}$ and, using the expression in \eqref{eq:steady_Psi},
\begin{equation*}
    \lim_{v\uparrow v_+}(v-\Psi(v,\lambda))=\beta z_{\max}>1/m,
\end{equation*}
where the strict inequality follows from Assumption \ref{asmp:bmz} because $z_{\max}\ge\E[Z\mid Z\ge1/m]$. On the other hand, concavity and \eqref{eq:MPK} imply $1/F_K(K(v),1)\le1/m$. Hence $H(v)<0$ near $v_+$. Since $H$ is continuous and strictly decreasing on $\cJ$, the intermediate value theorem yields a unique $v^*\in\cJ$ satisfying \eqref{eq:steady_Psi_v}. Lemma \ref{lem:steady_K} then yields the unique corresponding value $K^*\coloneqq K(v^*)$, and hence the unique admissible steady state $(K^*,v^*)$.
\end{proof}

\begin{proof}[Proof of Proposition \ref{prop:asymptotic}]
Lemma \ref{lem:steady_exist} proves the claims concerning admissible steady states. It remains to establish the claim concerning the growth rate $G$.
Letting $v=G/m$, \eqref{eq:steady} becomes $\Psi(v,\lambda)=0$. By Assumption \ref{asmp:Phi},
\begin{align*}
    \Psi(0,\lambda)&=-\beta\lambda\int_0^\infty z\diff\Phi(z)<0,\\
    \lim_{v\to\infty}\Psi(v,\lambda)&=\infty.
\end{align*}
Therefore, by the intermediate value theorem and the strict monotonicity in \eqref{eq:Psi_v}, there exists a unique $v>0$ such that $\Psi(v,\lambda)=0$. Equation \eqref{eq:Psi_threshold} implies that this root satisfies $v\gtreqless 1/m$ if and only if $\lambda\gtreqless\bar{\lambda}$. Because $G=mv$, there exists a unique $G>1$ solving \eqref{eq:steady} if and only if $\lambda>\bar{\lambda}$.
\end{proof}

\subsection{Proofs of Propositions \ref{prop:stable} and \ref{prop:exist}}

We use the following notation in both proofs: $f(K)\coloneqq F(K,1)$, $h(K)\coloneqq f(K)/K$,
\begin{align*}
    s(K)&\coloneqq\frac{F_X(K,1)}{f(K)}
    =1-\frac{Kf'(K)}{f(K)},     &I(v)&\coloneqq\int_v^\infty z\diff\Phi(z),\\
	b(v)&\coloneqq\beta(\lambda\Phi(v)+1-\lambda),
    &a(v)&\coloneqq\frac{\beta\lambda I(v)}{1-b(v)},
    \qquad q(v)\coloneqq\frac{b(v)}{1-b(v)}.
\end{align*}
Equations \eqref{eq:WPKz} and \eqref{eq:dynamics}, together with Euler's theorem, are equivalent to
\begin{subequations}\label{eq:local_reduced_system}
\begin{align}
    K_{t+1}&=a(\bar z_t)f(K_t),\qquad
    P_t=q(\bar z_t)f(K_t),\qquad
    W_t=\frac{f(K_t)}{1-b(\bar z_t)}, \label{eq:exist_reduced}\\
    q(\bar z_{t+1})
    &=\frac{\bar z_tq(\bar z_t)}{a(\bar z_t)}[1-s(K_{t+1})]-s(K_{t+1}). \label{eq:z_land_share}
\end{align}
\end{subequations}
Whenever $\Phi$ has a continuous density $\phi$ on the relevant neighborhood, differentiation gives
\begin{equation}
    a'(v)=\frac{\beta\lambda\phi(v)}{1-b(v)}[a(v)-v],
    \qquad
    q'(v)=\frac{\beta\lambda\phi(v)}{[1-b(v)]^2}. \label{eq:local_derivatives}
\end{equation}

\begin{proof}[Proof of Proposition \ref{prop:stable}]
By \eqref{eq:local_derivatives}, $q'(v^*)>0$, so \eqref{eq:z_land_share} locally defines $\bar z_{t+1}$ as a continuously differentiable function of $(K_t,\bar z_t)$. Thus the system defines a local map $T(K,v)=(A(K,v),B(K,v))$, where
\begin{equation*}
    A(K,v)=a(v)f(K),\qquad
    B(K,v)=q^{-1}\left(c(v)[1-s(A(K,v))]-s(A(K,v))\right),
\end{equation*}
and $c(v)\coloneqq vq(v)/a(v)$. At the steady state,
\begin{equation*}
    A_K=a(v^*)f'(K^*)=\frac{K^*f'(K^*)}{f(K^*)}
    =1-s(K^*)\in(0,1).
\end{equation*}
Equations \eqref{eq:Rt} and \eqref{eq:z_threshold} imply $f'(K^*)v^*>1$, and hence $v^*f(K^*)>K^*$. Therefore $v^*>a(v^*)$, so \eqref{eq:local_derivatives} gives $A_v=a'(v^*)f(K^*)<0$. Because $c=vq/a$ and $q'>0$, it follows that $c'(v^*)>0$.

Write $J=DT(K^*,v^*)$. To keep notation compact, in the next two displays all functions and derivatives are evaluated at $(K^*,v^*)$. Direct differentiation gives
\begin{equation}
    J=
    \begin{bmatrix}
        A_K & A_v \\
        -\dfrac{(1+c)s'A_K}{q'}
        & \dfrac{c'(1-s)-(1+c)s'A_v}{q'}
    \end{bmatrix}. \label{eq:local_J}
\end{equation}
Therefore the characteristic polynomial of $J$ is
\begin{align}
    \chi(\mu)&\coloneqq\det(\mu I-J)\notag \\
	&=\mu^2-\left[A_K+\frac{c'(1-s)-(1+c)s'A_v}{q'}\right]\mu
    +\frac{A_Kc'(1-s)}{q'}. \label{eq:local_charpoly}
\end{align}
In particular,
\begin{equation*}
    \chi(0)=\det J=A_Kc'(1-s)/q'>0.
\end{equation*}

It remains to determine the sign at one. Apply the implicit function theorem to $K=a(v)f(K)$ to obtain a local stationary-capital locus $K=K(v)$ with
\begin{equation*}
    K'(v^*)=\frac{A_v}{1-A_K}<0.
\end{equation*}
Let
\begin{equation*}
    r(v)\coloneqq q(v)-c(v)[1-s(K(v))]+s(K(v))
\end{equation*}
be the residual in the second steady-state equation along this locus. Substitution into \eqref{eq:local_J} gives
\begin{equation*}
    \chi(1)=\det(I-J)=\frac{1-A_K}{q'(v^*)}r'(v^*).
\end{equation*}
To sign this derivative, define
\begin{equation*}
    H(v)\coloneqq\frac{1}{f'(K(v))}-[v-\Psi(v,\lambda)].
\end{equation*}
Since $(v-\Psi(v,\lambda))'=b(v)>0$, concavity of $f$ implies
\begin{equation*}
    H'(v^*)=-\frac{f''(K^*)}{[f'(K^*)]^2}K'(v^*)-b(v^*)<0.
\end{equation*}
Using $a(v)=K(v)/f(K(v))$ and $v-\Psi(v,\lambda)=b(v)v+\beta\lambda I(v)$ gives
\begin{equation*}
    [1-b(v)]r(v)=f'(K(v))H(v).
\end{equation*}
At $v^*$, the second steady-state equation gives $r(v^*)=0$, while \eqref{eq:steady_Psi_v} gives $H(v^*)=0$. Differentiating this identity at $v^*$ therefore yields
\begin{equation*}
    r'(v^*)=\frac{f'(K^*)}{1-b(v^*)}H'(v^*)<0,
\end{equation*}
and hence $\chi(1)<0$.

Because \eqref{eq:local_charpoly} is a monic quadratic, $\chi(0)>0>\chi(1)$ implies that it has one root $\mu_s\in(0,1)$. Moreover, $\chi(1)<0$ and $\chi(\mu)\to\infty$ as $\mu\to\infty$, so it has another root $\mu_u>1$. Thus $(K^*,v^*)$ is a hyperbolic saddle with a one-dimensional stable direction. Since $\chi(0)=\det J>0$, $T$ is a local diffeomorphism. The stable direction is not vertical because $A_v<0$, so the local stable manifold theorem \citep[Theorem~8.9]{TodaEME} yields a local stable manifold that can be written as a graph $v=g(K)$. For every $K_0$ sufficiently close to $K^*$, choosing $\bar z_0=g(K_0)$ generates a locally unique path converging to $(K^*,v^*)$. Shrinking the neighborhood preserves positivity and admissibility, and Lemma \ref{lem:dynamics} recovers the remaining equilibrium variables. By \eqref{eq:exist_reduced} and continuity, wealth, land prices, and land rents converge to positive limits, so the equilibrium exhibits asymptotically balanced growth.
\end{proof}

We next collect two consequences of the elasticity bound for high capital. The first will also be used in the proof of Theorem \ref{thm:land_bubble}.

\begin{lem}[High-capital production bounds]\label{lem:MRT}
Suppose Assumption \ref{asmp:F} holds. Then:
\begin{enumerate}
    \item\label{item:MRT1} Let $\ubar{K}>0$ and $\ubar{\sigma}>0$, and suppose that
    $\sigma(K,1)\ge\ubar{\sigma}$ for all $K\ge\ubar{K}$. Then
    \begin{equation}
        \frac{F_X}{F_K}(K,1)
        \le \frac{F_X}{F_K}(\ubar{K},1)
        (K/\ubar{K})^{1/\ubar{\sigma}}
        \quad\text{for all }K\ge\ubar{K}. \label{eq:MRT_ub}
    \end{equation}
    \item\label{item:MRT2} If Assumption \ref{asmp:ES} additionally holds, there exist constants
    $C,\eta>0$ and $\ubar{K}>0$ such that, for all $K\ge\ubar{K}$,
    \begin{align}
        &0<s(K)\le CK^{-\eta}, &&
        \abs{Ks'(K)}\le CK^{-\eta}, \label{eq:s_decay_bound} \\
        &0<h(K)-m\le CK^{-\eta}, &&
        \abs*{\frac{\diff\log h(K)}{\diff\log K}}
        =s(K)\le CK^{-\eta}. \label{eq:h_decay_bound}
    \end{align}
\end{enumerate}
\end{lem}

\begin{proof}
\ref{item:MRT1} Because $F_K/F_X$ is homogeneous of degree zero, it depends only on the factor ratio $K/X$. Hence the definition of the elasticity in \eqref{eq:sigma} implies
\begin{equation}
    \frac{\diff\log (F_X/F_K)(K,1)}{\diff\log K}
    =\frac{1}{\sigma(K,1)}. \label{eq:MRT_derivative}
\end{equation}
Under the hypothesis, the right-hand side is at most $1/\ubar{\sigma}$. Integrating from $\ubar{K}$ to $K$ and exponentiating yields \eqref{eq:MRT_ub}.

\medskip
\noindent \ref{item:MRT2} Choose $\ubar{\sigma}\in (1,\liminf_{K\to\infty}\sigma(K,1))$. Part \ref{item:MRT1} implies that for some $C_0>0$, we have
$(F_X/F_K)(K,1)\le C_0K^{1/\ubar{\sigma}}$
for all sufficiently large $K$. Moreover, \eqref{eq:MPK} implies
\begin{equation*}
    \frac{F_K(K,1)}{h(K)}
    =\frac{F_K(K,1)}{f(K)/K}\to1.
\end{equation*}
It follows that, for some $C>0$ and all sufficiently large $K$,
\begin{equation*}
    s(K)=\frac{1}{K}\frac{F_X}{F_K}(K,1)
    \frac{F_K(K,1)}{h(K)}
    \le CK^{1/\ubar{\sigma}-1}.
\end{equation*}
Thus the first bound in \eqref{eq:s_decay_bound} holds with
$\eta=1-1/\ubar{\sigma}>0$.

We next control the derivative of the land share. Euler's theorem and the definition of $s$ imply
\begin{equation*}
    \frac{F_K}{F_X}(K,1)
    =\frac{f'(K)}{s(K)f(K)}
    =\frac{1-s(K)}{Ks(K)}.
\end{equation*}
Taking the negative of \eqref{eq:MRT_derivative}, and also taking logarithms of the preceding expression for $F_K/F_X$ and differentiating with respect to $\log K$, yields
\begin{align*}
    -\frac{1}{\sigma(K,1)}
    &=-\frac{Ks'(K)}{1-s(K)}-\frac{Ks'(K)}{s(K)}-1\\
    &=-\frac{1}{1-s(K)}\frac{\diff\log s(K)}{\diff\log K}-1.
\end{align*}
Rearranging, we obtain
\begin{equation}
    -\frac{\diff\log s(K)}{\diff\log K}
    =\left(1-\frac{1}{\sigma(K,1)}\right)(1-s(K)). \label{eq:land_share_decay}
\end{equation}
For all sufficiently large $K$, we have $\sigma(K,1)>1$ and $0<s(K)<1$.
It follows from \eqref{eq:land_share_decay} that
\begin{equation*}
    \abs{Ks'(K)}
    =s(K)\abs*{\frac{\diff\log s(K)}{\diff\log K}}
    \le s(K)\le CK^{-\eta},
\end{equation*}
which proves the second bound in \eqref{eq:s_decay_bound}.

Finally, Euler's theorem implies
\begin{equation*}
    h'(K)=\frac{Kf'(K)-f(K)}{K^2}
    =-\frac{F_X(K,1)}{K^2}
    =-\frac{s(K)h(K)}{K}.
\end{equation*}
Since $h(K)\downarrow m$ as $K\to\infty$, for all sufficiently large $K$ we have
\begin{align*}
    h(K)-m
    &=-\int_K^\infty h'(u)\diff u
      =\int_K^\infty\frac{s(u)h(u)}{u}\diff u\\
    &\le C\int_K^\infty u^{-1-\eta}\diff u
      \le CK^{-\eta},
\end{align*}
after increasing $C$ if necessary. The identity
$\diff\log h(K)/\diff\log K=-s(K)$ proves the remaining claim in \eqref{eq:h_decay_bound}.
\end{proof}

\begin{proof}[Proof of Proposition \ref{prop:exist}]
Fix constants $C,\eta,\ubar{K}$ satisfying
\eqref{eq:s_decay_bound}--\eqref{eq:h_decay_bound}.
Because $\Psi(\bar z,\lambda)=0$, \eqref{eq:steady_Psi} implies
\begin{equation}
    \beta\lambda I(\bar z)=(1-b(\bar z))\bar z,
    \qquad a(\bar z)=\bar z,
    \qquad G=ma(\bar z)=m\bar z. \label{eq:exist_root}
\end{equation}
By the assumed admissibility of $\bar z$, we have $b(\bar z)\in(0,\beta)$. Equation \eqref{eq:local_derivatives} gives $q'(\bar z)>0$, so $q$ has a continuously differentiable local inverse; together with \eqref{eq:exist_root}, it also gives $a'(\bar z)=0$. Define the limiting threshold map
\begin{equation*}
    T(v)\coloneqq q^{-1}(vq(v)/a(v)).
\end{equation*}
Equation \eqref{eq:exist_root} implies $T(\bar z)=\bar z$, and
\begin{equation*}
    \Lambda\coloneqq T'(\bar z)
    =1+\frac{b(\bar z)(1-b(\bar z))}
    {\beta\lambda\bar z\phi(\bar z)}>1.
\end{equation*}

We construct the equilibrium by a contraction argument. Define the detrended variables
\begin{equation}
    \delta_t\coloneqq\bar z_t-\bar z,
    \qquad
    x_t\coloneqq\log\frac{K_t}{K_0G^t}. \label{eq:xdelta}
\end{equation}
For $\delta$ and $y$ near zero, set
\begin{align*}
    \alpha(\delta)&\coloneqq\log\frac{a(\bar z+\delta)}{a(\bar z)}, \qquad \epsilon(K)\coloneqq\log\frac{h(K)}{m},\\
    N(\delta)&\coloneqq T(\bar z+\delta)-\bar z-\Lambda\delta,\\
    V(\delta,y)&\coloneqq q^{-1}\left(
        \frac{(\bar z+\delta)q(\bar z+\delta)}{a(\bar z+\delta)}(1-y)-y
        \right)-T(\bar z+\delta).
\end{align*}
Equations \eqref{eq:exist_reduced} and \eqref{eq:z_land_share} are equivalent to
\begin{subequations}\label{eq:xdelta_dynamics}
\begin{align}
    x_{t+1}-x_t
    &=\alpha(\delta_t)+\epsilon(K_0G^t\e^{x_t}), \label{eq:x_dynamics}\\
    \delta_{t+1}
    &=\Lambda\delta_t+N(\delta_t)
      +V\left(\delta_t,s(K_0G^{t+1}\e^{x_{t+1}})\right). \label{eq:delta_dynamics}
\end{align}
\end{subequations}
The preceding definitions and $a'(\bar z)=0$ give
\begin{equation}
    \alpha(0)=\alpha'(0)=N(0)=N'(0)=0,
    \qquad V(\delta,0)=0. \label{eq:small_terms}
\end{equation}
Furthermore, $V_y$ is locally bounded and $V_\delta(\delta,y)\to0$ as $(\delta,y)\to(0,0)$.

Choose $\tau\in(G^{-\eta},1)$. On pairs of sequences $(x,\delta)=\set{(x_t,\delta_t)}_{t=0}^\infty$, use the norm
\begin{equation*}
    \norm{(x,\delta)}
    \coloneqq\sup_{t\ge0}\abs{x_t}
    +\sup_{t\ge0}\tau^{-t}\abs{\delta_t}.
\end{equation*}
The boundary conditions $x_0=0$ and $\delta_t\to0$ allow us to write \eqref{eq:xdelta_dynamics} as the fixed-point equations $x_0=0$,
\begin{subequations}\label{eq:fixed_point_sequences}
\begin{align}
    x_t&=\sum_{j=0}^{t-1}
    \left[\alpha(\delta_j)+\epsilon(K_0G^j\e^{x_j})\right], \label{eq:fixed_x}\\
    \delta_t&=-\sum_{j=t}^\infty\Lambda^{t-j-1}
    \left[N(\delta_j)+V\left(\delta_j,s(K_0G^{j+1}\e^{x_{j+1}})\right)\right]. \label{eq:fixed_delta}
\end{align}
\end{subequations}
Conversely, any solution to \eqref{eq:fixed_point_sequences} satisfies \eqref{eq:xdelta_dynamics} and $\delta_t\to0$.

For completeness, we verify the contraction property. Fix a sufficiently large constant $M$ and consider the closed ball
\begin{equation}
    \sup_{t\ge0}\abs{x_t}\le MK_0^{-\eta},
    \qquad
    \sup_{t\ge0}\tau^{-t}\abs{\delta_t}\le MK_0^{-\eta}. \label{eq:sequence_ball}
\end{equation}
For sufficiently large $K_0$, every pair in this ball satisfies $K_0G^t\e^{x_t}\ge \frac{1}{2}K_0G^t$. Therefore \eqref{eq:s_decay_bound}--\eqref{eq:h_decay_bound} and $G^{-\eta}<\tau$ imply, uniformly on the ball,
\begin{equation}
    s(K_0G^t\e^{x_t})+
    \abs{\epsilon(K_0G^t\e^{x_t})}
    \le CK_0^{-\eta}\tau^t. \label{eq:forcing_bound}
\end{equation}
    The derivatives of these two expressions with respect to $x_t$ satisfy the same bound by \eqref{eq:s_decay_bound}--\eqref{eq:h_decay_bound}. By \eqref{eq:small_terms}, the suprema of $\abs{\alpha'(\delta)}$, $\abs{N'(\delta)}$, and $\abs{V_\delta(\delta,y)}$ on the relevant ranges converge to zero as $K_0\to\infty$, whereas $V_y$ remains bounded. Finally,
\begin{equation*}
    \sum_{j=0}^\infty\tau^j=\frac{1}{1-\tau}<\infty,
    \qquad
    \sum_{j=t}^\infty\Lambda^{t-j-1}\tau^j
    =\frac{\tau^t}{\Lambda-\tau}.
\end{equation*}
These estimates and the mean value theorem show first that the right-hand side of \eqref{eq:fixed_point_sequences} maps \eqref{eq:sequence_ball} into itself when $M$ is large and $K_0$ is sufficiently large, and second that its Lipschitz constant converges to zero as $K_0\to\infty$. It is therefore a contraction for all sufficiently large $K_0$. The contraction mapping theorem yields a unique local solution $(x,\delta)$, and hence, through \eqref{eq:xdelta}, a locally unique equilibrium path satisfying
\begin{equation}
    \abs{\bar z_t-\bar z}\le MK_0^{-\eta}\tau^t. \label{eq:z_convergence_rate}
\end{equation}

It remains to establish the level limits. Equations \eqref{eq:forcing_bound}, \eqref{eq:z_convergence_rate}, and $\alpha(0)=\alpha'(0)=0$ imply that the increments on the right-hand side of \eqref{eq:x_dynamics} are absolutely summable. Hence $x_t\to x_\infty$ for some finite $x_\infty$, and
\begin{equation*}
    K_t/G^t\to k\coloneqq K_0\e^{x_\infty}>0.
\end{equation*}
Using \eqref{eq:exist_reduced}, $h(K_t)\to m$, and $\bar z_t\to\bar z$, we finally obtain
\begin{equation*}
    P_t/G^t\to p\coloneqq q(\bar z)mk>0,
    \qquad
    W_t/G^t\to w\coloneqq\frac{mk}{1-b(\bar z)}>0. \qedhere
\end{equation*}
\end{proof}

\subsection{Proof of Theorem \ref{thm:land_bubble}}
\ref{item:land_bubble_balanced} 
Suppose $\lambda<\bar{\lambda}$ and consider any equilibrium converging to the steady state. Then by definition $r_t/P_t$ converges to a finite positive number, so $\sum_{t=1}^\infty r_t/P_t=\infty$. By the Bubble Characterization Lemma in \eqref{eq:bubble_characterization}, we have $P_t=V_t$.

\medskip
\noindent \ref{item:land_bubble_unbalanced}
Next, suppose $\lambda>\bar{\lambda}$ and consider the equilibrium in Proposition \ref{prop:exist}. By Assumption \ref{asmp:ES}, we may choose $\ubar{\sigma}>1$ and $\ubar{K}>0$ such that $\sigma(K,1)\ge\ubar{\sigma}$ for all $K\ge\ubar{K}$. Set $\nu\coloneqq1/\ubar{\sigma}\in(0,1)$. Proposition \ref{prop:exist} gives $K_t/G^t\to k>0$, so $K_t\ge\ubar{K}$ for all sufficiently large $t$. Lemma \ref{lem:MRT}\ref{item:MRT1} and \eqref{eq:MPK} then imply
\begin{equation*}
    r_t=F_X(K_t,1)
    \le F_K(K_t,1)\frac{F_X}{F_K}(\ubar{K},1)
    (K_t/\ubar{K})^\nu
    =O(G^{\nu t}).
\end{equation*}
Because $P_t/G^t\to p>0$, it follows that $r_t/P_t=O(G^{-(1-\nu)t})$. This sequence is summable because $\nu<1$. By the Bubble Characterization Lemma in \eqref{eq:bubble_characterization}, we have $P_t>V_t$ for all $t$. \hfill \qedsymbol

\section*{Declaration of generative AI and AI-assisted technologies in the manuscript preparation process}

During the preparation of this work, the authors used ChatGPT (OpenAI) to assist with language editing, manuscript organization, literature searches, and checks of exposition and internal consistency. The author reviewed and edited all AI-assisted output and takes full responsibility for the content of the published article.

\printbibliography

\clearpage

\begin{center}
    {\Large \textbf{Online Appendix (Not for publication)}}
\end{center}

\section{Implementation of land trading}\label{sec:land_trading}

In the main text, $a_{it}\in\R$ denotes the date-$t$ market value of agent $i$'s total position in risk-free claims. In \S\ref{sec:simple}, $a_{it}=P_t\theta_{it}+n_{it}$ combines land and the bond traded with the financial sector; in \S\ref{sec:general}, there is no such sector and $N_t=0$. Land trading can equivalently be implemented through a pass-through real estate investment trust (REIT).

Let a perfectly competitive REIT hold the unit aggregate supply of land and issue one unit of shares. Denote the ex-dividend price per share by $Q_t$ and agent $i$'s share holdings by $\theta_{it}\in\R$. One share purchased at date $t$ pays the next ex-dividend price plus the land rent, $Q_{t+1}+F_X(K_{t+1},1)$, at date $t+1$. Because we restrict attention to deterministic aggregate equilibria, the share is risk-free and has gross return
\begin{equation}
    R_t=\frac{Q_{t+1}+F_X(K_{t+1},1)}{Q_t}. \label{eq:REIT_return}
\end{equation}
The REIT is a one-for-one pass-through claim to land, so competition and zero profit imply $Q_t=P_t$. Equation \eqref{eq:REIT_return} is therefore exactly the return equation \eqref{eq:Rt}.

Define the market value of agent $i$'s REIT position by
\begin{equation*}
    a_{it}\coloneqq Q_t\theta_{it}.
\end{equation*}
This position costs $a_{it}$ at date $t$ and pays $R_ta_{it}$ at date $t+1$, exactly as in the individual budget constraints in the main text. REIT share-market clearing gives
\begin{equation}
    \int_I \theta_{it}\diff i=1
    \quad\Longrightarrow\quad
    \int_I a_{it}\diff i=Q_t=P_t. \label{eq:REIT_clearing}
\end{equation}
Integrating the optimal asset positions in \eqref{eq:asset_allocation}, with the time-varying threshold $\bar{z}_t$, and using \eqref{eq:REIT_clearing} yields
\begin{equation*}
    P_t=\beta\bigl(\lambda\Phi(\bar{z}_t)+1-\lambda\bigr)W_t,
\end{equation*}
which is \eqref{eq:Pt1} with $N_t=0$. Moreover, the aggregate payoff from date-$(t-1)$ REIT positions is
\begin{equation*}
    R_{t-1}\int_I a_{i,t-1}\diff i
    =R_{t-1}P_{t-1}=P_t+F_X(K_t,1).
\end{equation*}
Adding aggregate capital income and applying Euler's theorem gives
\begin{equation*}
    W_t=F(K_t,1)+P_t,
\end{equation*}
as in \eqref{eq:Wt}. Hence the REIT implementation reproduces the individual and aggregate equilibrium conditions in the main text.

Alternatively, one may rule out short positions in REIT shares by imposing $\theta_{it}\ge0$ and introducing a one-period risk-free bond in zero net supply. Let $b_{it}\in\R$ be the date-$t$ market value of agent $i$'s bond position, which pays $R_tb_{it}$ at date $t+1$, and define $a_{it}=Q_t\theta_{it}+b_{it}$. No arbitrage makes the bond and the REIT perfect substitutes, so only $a_{it}$ matters for individual saving, borrowing, and the leverage constraint. Because $\int_I\theta_{it}\diff i=1$ and $\int_I b_{it}\diff i=0$, aggregate market clearing again gives $\int_I a_{it}\diff i=P_t$. This formulation therefore yields the same aggregate equilibrium system without requiring agents to short REIT shares. In the tractable benchmark, replacing the zero-net-supply bond $b_{it}$ with the position $n_{it}$ vis-\`a-vis the financial sector gives $a_{it}=Q_t\theta_{it}+n_{it}$ and $\int_Ia_{it}\diff i=P_t+N_t$, exactly as in \eqref{eq:Pt1}.

\section{Solving the numerical illustration}\label{sec:num}

This appendix explains how we solve the numerical illustration in \S\ref{sec:discuss}. We use the general model of \S\ref{sec:general} with the exponential productivity distribution \eqref{eq:exp}.

\subsection{Derivation of formulas used in the main text}

Using integration by parts, the expected excess return \eqref{eq:pi} is
\begin{align*}
    \pi(R)&=\int_{R/m}^\infty (mz-R)\diff \Phi(z)=\int_{R/m}^\infty(R-mz)\diff (1-\Phi(z))\\
    &=[(R-mz)(1-\Phi(z))]_{R/m}^\infty+\int_{R/m}^\infty m(1-\Phi(z))\diff z\\
    &=m\int_{R/m}^\infty \e^{-\gamma z}\diff z=\frac{m}{\gamma}\e^{-\frac{\gamma R}{m}}.
\end{align*}
Setting $R=1$ in this expression and substituting it into \eqref{eq:lambda_threshold} yields \eqref{eq:lambda_exmp}. A second integration by parts gives
\begin{align*}
    \int_{\bar{z}}^\infty z\diff\Phi(z)&=\int_{\bar{z}}^\infty z\gamma\e^{-\gamma z}\diff z=\int_{\bar{z}}^\infty z(-\e^{-\gamma z})'\diff z\\
    &=\left[-z\e^{-\gamma z}\right]_{\bar{z}}^\infty+\int_{\bar{z}}^\infty \e^{-\gamma z}\diff z=(\bar{z}+1/\gamma)\e^{-\gamma \bar{z}},
\end{align*}
which proves \eqref{eq:zint}. The equilibrium conditions used in the proof of Theorem \ref{thm:trend} relate leverage and the interest rate according to
\begin{equation*}
    \lambda=
    \begin{cases*}
        \dfrac{1-\beta R}{\beta\pi(R)}=\dfrac{\gamma(1-\beta R)}{\beta m}\e^{\gamma R/m} & in balanced growth,\\
        \dfrac{(1-\beta)R}{\beta\pi(R)}=\dfrac{\gamma(1-\beta)}{\beta m}R\e^{\gamma R/m} & in unbalanced growth.
    \end{cases*}
\end{equation*}
We use these formulas to plot Figure~\ref{fig:rate}.

\subsection{Long-run values}

We first compute the limiting productivity cutoff in each long-run regime. If $\lambda>\bar{\lambda}$, imposing $\bar{z}=G/m$ in \eqref{eq:steady} gives
\begin{equation*}
    (1-\beta)\bar{z}=\frac{\beta\lambda}{\gamma}\e^{-\gamma\bar{z}}.
\end{equation*}
We solve this equation numerically for the limiting cutoff $\bar{z}$ and obtain the candidate growth rate $G=m\bar{z}$, subject to the admissibility condition in Proposition \ref{prop:exist}.

If $\lambda<\bar{\lambda}$, we proceed in two steps. First, fix an arbitrary $\bar{z}>0$, impose $\bar{z}_t=\bar{z}$ and $K_{t+1}=K_t=K$ in \eqref{eq:dynamics_K1}, and solve for $K$. Let
\begin{equation*}
    C=\frac{\beta\lambda(\bar{z}+1/\gamma)\e^{-\gamma\bar{z}}}{1-\beta+\beta\lambda\e^{-\gamma\bar{z}}}.
\end{equation*}
Solving directly for the steady-state capital stock under the CES production function \eqref{eq:CES} gives
\begin{equation*}
    K(\bar{z})\coloneqq\begin{cases*}
        \left(\frac{\left(\frac{1-C(1-\delta)}{CA}\right)^{1-\rho}-\alpha}{1-\alpha}\right)^\frac{1}{\rho-1} & if $\rho\neq 1$,\\
        \left(\frac{1-C(1-\delta)}{CA}\right)^\frac{1}{\alpha-1} & if $\rho=1$.
    \end{cases*}
\end{equation*}
For $K(\bar{z})$ to be well defined and strictly positive, we require
\begin{equation*}
    C(1-\delta)<1
    \quad\text{and}\quad
    \left(\frac{1-C(1-\delta)}{CA}\right)^{1-\rho}>\alpha.
\end{equation*}
If either inequality is violated, no finite positive solution exists, and we discard that candidate value of $\bar{z}$. Second, impose $K_{t+1}=K_t=K(\bar{z})$ and $\bar{z}_{t+1}=\bar{z}_t=\bar{z}$ in \eqref{eq:dynamics_z} and solve for $\bar{z}$.

\subsection{Transition dynamics}

To solve for the transition dynamics, choose a large terminal horizon $T$ and initialize a guess $\set{\bar{z}_t}_{t=0}^T$, fixing $\bar{z}_T$ at the long-run value computed above. Given initial aggregate capital $K_0$, generate $\set{K_t}_{t=0}^T$ forward using \eqref{eq:dynamics_K1}. Then use \eqref{eq:dynamics_z} to update the cutoff sequence backward according to
\begin{equation*}
    \bar{z}_t^\mathrm{new}=\frac{P(K_{t+1},\bar{z}_{t+1})+F_X(K_{t+1},1)}{F_K(K_{t+1},1)P(K_t,\bar{z}_t)},
    \qquad t=0,\dots,T-1.
\end{equation*}
We find $\set{\bar{z}_t}_{t=0}^T$ by minimizing the equilibrium error
\begin{equation*}
    \sum_{t=0}^{T-1}(\bar{z}_t^\mathrm{new}-\bar{z}_t)^2.
\end{equation*}
To reduce the dimensionality, we parameterize the log deviation $y_t=\log(\bar{z}_t/\bar{z}_T)$ at $J=10$ nodes on a logarithmically spaced time grid, impose $y_T=0$, and use cubic spline interpolation to construct the full sequence $\set{\bar{z}_t}_{t=0}^T$. For Figure~\ref{fig:dynamics}, we apply this procedure separately after each unanticipated leverage change and join the resulting deterministic transition paths at $t=0$ and $t=10$.

\end{document}